\documentclass[aps,amsmath,amssymb, pra,superscriptaddress,reprint]{revtex4-2}

\usepackage{hyperref}
\usepackage{graphicx}
\usepackage{subfigure}
\usepackage{bm}

\begin{document}

\title{Electron-loss-to-continuum cusp in collisions of U$^\mathbf{89+}$ with N$_\mathbf{2}$ and Xe}

\author{P.-M.~Hillenbrand}
\affiliation{Institut f\"ur Kernphysik, Goethe-Universit\"at, 60438 Frankfurt, Germany}
\affiliation{GSI Helmholtzzentrum f\"ur Schwerionenforschung, 64291 Darmstadt, Germany}

\author{K.~N.~Lyashchenko}
\affiliation{Institute of Modern Physics, Chinese Academy of Sciences, Lanzhou 730000, China}

\author{S.~Hagmann}
\affiliation{GSI Helmholtzzentrum f\"ur Schwerionenforschung, 64291 Darmstadt, Germany}

\author{O.~Yu.~Andreev}
\affiliation{Department of Physics, St.~Petersburg State University, St. Petersburg, 199034, Russia}
\affiliation{Petersburg Nuclear Physics Institute named by B.P.~Konstantinov of National Research Centre “Kurchatov Institute”, Gatchina, Leningrad District, 188300, Russia}

\author{D.~Bana\'{s}}
\affiliation{Institute of Physics, Jan Kochanowski University, 25-406 Kielce, Poland}

\author{E.~P.~Benis}
\affiliation{Department of Physics, University of Ioannina, 45110 Ioannina, Greece}

\author{A.~I.~Bondarev}
\affiliation{Center for Advanced Studies, Peter the Great St.~Petersburg Polytechnic University, St.~Petersburg, 195251, Russia}

\author{C.~Brandau}
\affiliation{GSI Helmholtzzentrum f\"ur Schwerionenforschung, 64291 Darmstadt, Germany}
\affiliation{I.~Physikalisches Institut, Justus-Liebig-Universit\"at, 35392 Giessen, Germany}

\author{E.~De~Filippo}
\affiliation{Istituto Nazionale di Fisica Nucleare, Sezione di Catania, 95123 Catania, Italy}

\author{O.~Forstner}
\affiliation{GSI Helmholtzzentrum f\"ur Schwerionenforschung, 64291 Darmstadt, Germany}
\affiliation{Institut f\"ur Optik und Quantenelektronik, Friedrich-Schiller-Universit\"at, 07743 Jena, Germany}

\author{J.~Glorius}
\affiliation{GSI Helmholtzzentrum f\"ur Schwerionenforschung, 64291 Darmstadt, Germany}

\author{R.~E.~Grisenti}
\affiliation{Institut f\"ur Kernphysik, Goethe-Universit\"at, 60438 Frankfurt, Germany}
\affiliation{GSI Helmholtzzentrum f\"ur Schwerionenforschung, 64291 Darmstadt, Germany}

\author{A.~Gumberidze}
\affiliation{GSI Helmholtzzentrum f\"ur Schwerionenforschung, 64291 Darmstadt, Germany}

\author{D.~L.~Guo}
\affiliation{Institute of Modern Physics, Chinese Academy of Sciences, Lanzhou 730000, China}


\author{M.~O.~Herdrich}
\affiliation{Institut f\"ur Optik und Quantenelektronik, Friedrich-Schiller-Universit\"at, 07743 Jena, Germany}
\affiliation{Helmholtz-Institut Jena, 07743 Jena, Germany}

\author{M.~Lestinsky}
\affiliation{GSI Helmholtzzentrum f\"ur Schwerionenforschung, 64291 Darmstadt, Germany}

\author{Yu.~A.~Litvinov}
\affiliation{GSI Helmholtzzentrum f\"ur Schwerionenforschung, 64291 Darmstadt, Germany}

\author{E.~V.~Pagano}
\affiliation{Istituto Nazionale di Fisica Nucleare, Laboratori Nazionali del Sud, 95123, Catania, Italy}

\author{N.~Petridis}
\affiliation{GSI Helmholtzzentrum f\"ur Schwerionenforschung, 64291 Darmstadt, Germany}

\author{M.~S.~Sanjari}
\affiliation{GSI Helmholtzzentrum f\"ur Schwerionenforschung, 64291 Darmstadt, Germany}
\affiliation{Aachen University of Applied Sciences, 52066 Aachen, Germany}

\author{D.~Schury}
\affiliation{GSI Helmholtzzentrum f\"ur Schwerionenforschung, 64291 Darmstadt, Germany}
\affiliation{I.~Physikalisches Institut, Justus-Liebig-Universit\"at, 35392 Giessen, Germany}

\author{U.~Spillmann}
\affiliation{GSI Helmholtzzentrum f\"ur Schwerionenforschung, 64291 Darmstadt, Germany}

\author{S.~Trotsenko}
\affiliation{GSI Helmholtzzentrum f\"ur Schwerionenforschung, 64291 Darmstadt, Germany}

\author{M.~Vockert}
\affiliation{Institut f\"ur Optik und Quantenelektronik, Friedrich-Schiller-Universit\"at, 07743 Jena, Germany}
\affiliation{Helmholtz-Institut Jena, 07743 Jena, Germany}

\author{A.~B.~Voitkiv}
\affiliation{Institut f\"ur Theoretische Physik I, Heinrich Heine Universität, 40225 D\"usseldorf, Germany}

\author{G.~Weber}
\affiliation{Helmholtz-Institut Jena, 07743 Jena, Germany}

\author{Th.~St\"ohlker}
\affiliation{GSI Helmholtzzentrum f\"ur Schwerionenforschung, 64291 Darmstadt, Germany}
\affiliation{Institut f\"ur Optik und Quantenelektronik, Friedrich-Schiller-Universit\"at, 07743 Jena, Germany}
\affiliation{Helmholtz-Institut Jena, 07743 Jena, Germany}

\date{\today}

\begin{abstract}
We study the electron-loss-to-continuum (ELC) cusp experimentally and theoretically by comparing the ionization of U$^{89+}$ projectiles in collisions with N$_2$ and Xe targets, at a beam energy of 75.91~MeV/u. The coincidence measurement between the singly ionized projectile and the energy of the emitted electron is used to compare the shape of the ELC cusp at weak  and strong perturbations. A significant  energy shift for the centroid of the electron cusp is observed for the heavy target of Xe as compared to the light target of N$_2$. Our results provide a stringent test for fully relativistic calculations of double-differential cross sections performed in the first-order approximation and in the continuum-distorted-wave approach.

\end{abstract}

\maketitle

\section{\label{sec:introduction}Introduction}

The energy distribution of emitted electrons is a prominent observable to study dynamical processes occurring in collisions of highly charged projectile ions with neutral targets \cite{stolterfoht_mechanisms_1974,stolterfoht_electron_1997}. In these collisions, electrons emitted into the projectile continuum appear as well-known `cusp electrons' in the laboratory frame, i.e., electrons ejected at zero degrees with respect to the projectile beam, with velocities comparable to the projectile velocity 
\cite{rudd_energy_1966,macek_theory_1970,crooks_experimental_1970,dettmann_charge_1974,macek_evidence_1981,breinig_experiments_1982}. In particular, the electron-loss-to-continuum (ELC) cusp, observed as coincidence events between the emitted electron and the ionized projectile, provides a highly sensitive tool to probe the energy and angular differential cross section of projectile ionization \cite{drepper_doubly_1976,briggs_asymptotic_1978,briggs_structure_1980,burgdorfer_calculation_1983,elston_observation_1985}. For nonrelativistic collision systems, asymmetries occurring in the ELC cusp were attributed to higher-order effects  \cite{oswald_higher-order_1989,atan_multipole_1990,jakubassa-amundsen_second_1990,gulyas_cusp-shape_1992,jakubassa-amundsen_strong_1993,hillenbrand_strong_2016}. The advantage of the electron cusp is that the transformation of the continuum electron phase space from the projectile frame to the target frame results in a sampling of the major characteristic features along the energy axis of forward-emitted electrons \cite{hillenbrand_electron-loss--continuum_2014,hillenbrand_radiative_2020}. This sampling eliminates the need to select arguable cuts of the continuum electron phase space for comparison of experiment and theory.

The accurate description for the inner-shell ionization of heavy ions through the collision with an atom requires the application of relativistic wave functions for the initial bound state and the final continuum state of the electron in the field of the projectile ion \cite{voitkiv_relativistic_2007}. The ionization of a projectile ion by a target atom can be characterized by the perturbation parameter $\nu_\mathrm{t}=\alpha Z_\mathrm{t}/v_\mathrm{p}$, with
$\alpha$ being the fine-structure constant, $Z_\mathrm{t}$ the atomic number of the target atom, and $v_\mathrm{p}$ the projectile velocity expressed in units of the speed of light. For $\nu_\mathrm{t}\ll1$, the ionization process is appropriately described by taking into account the first-order interaction between the projectile's electron and the target nucleus~\cite{voitkiv_relativistic_2008}. 
Fully relativistic first-order approaches are meanwhile well established \cite{momberger_angular_1989,voitkiv_plane-wave_2000,surzhykov_electron_2005,lyashchenko_electron_2018,bondarev_calculations_2020}. In those calculations, the magnitude of the double-differential cross section (DDCS) is proportional to $Z_\mathrm{t}^2$, and its shape does not depend on $Z_\mathrm{t}$, yielding a symmetric electron cusp shape in momentum space. For such a case of a heavy few-electron projectile, U$^{88+}(1s^22s^2)$, ionized by a N$_2$ target, we have recently reported a good agreement between the experimental and theoretical differential cross sections for the ELC cusp \cite{hillenbrand_electron-loss--continuum_2014}.

In the nonrelativistic regime, the continuum-distorted-wave (CDW) approach is well established for calculating electron continua for a wide range of perturbation parameters \cite{belkic_quantum_1978,belkic_electron_1979,crothers_ionisation_1983,maidagan_symmetric_1984,deco_symmetric_1986,fainstein_symmetric_1987,martinez_second-order_1990,fainstein_two-centre_1991,dewangan_charge_1994}. However, combining the CDW approach with fully relativistic wave functions has been a major challenge~\cite{toshima_distorted-wave_1990,crothers_relativistic_2000}. Considerable progress in this field was obtained in works  \cite{voitkiv_three-body_2007,voitkiv_electron_2007}, in which the total cross section of electron loss from heavy ions during their collisions with various nuclei was investigated. In the present study we performed relativistic CDW calculations for the DDCS of the electron cusp and conducted corresponding measurements to provide stringent tests for our theoretical predictions. This is done by comparing a collision system with a light target, which is appropriately described by a weak perturbation and a first-order approximation, to a collision system with a heavy target corresponding to a strong perturbation, which requires a more sophisticated description such as CDW theory.

To this end, we focus on comparing collisions of Li-like U$^{89+}\left(1s^22s\right)$ projectiles ionized by a light nitrogen target,
\begin{equation}\label{eq:Ntarget}
	{\rm U}^{89+} + {\rm N} \rightarrow {\rm U}^{90+}+{\rm N}^* + {\rm e}^-(E_\mathrm{e},\vartheta_\mathrm{e}),
\end{equation}
with the ionization by a heavy xenon target,
\begin{equation}\label{eq:Xetarget}
	{\rm U}^{89+} + {\rm Xe} \rightarrow {\rm U}^{90+}+{\rm Xe}^* + {\rm e}^-(E_\mathrm{e},\vartheta_\mathrm{e}).
\end{equation}
Our observable is the energy distribution of the emitted electrons, $E_\mathrm{e}$, detected at a polar angle of $\vartheta_\mathrm{e}= 0^\circ$ with respect to the projectile beam. For these collision systems, the molecular character of the N$_2$ target may be ignored, since the molecular binding energy is negligible compared to the collision energy and the target molecules are aligned randomly with respect to the collision axis. Furthermore, the target atom is expected to be multiply ionized in the collision. This, however, is not in the focus of the current study, therefore we mark the unresolved charge state of the outgoing target atom in Eqs.~(\ref{eq:Ntarget}) and (\ref{eq:Xetarget}) by a star symbol. At the applied collision energy of 75.91~MeV/u, the nitrogen target constitutes a weak perturbation of $\nu_\mathrm{t}=0.13$, while the xenon target leads to a strong perturbation of $\nu_\mathrm{t}=1.04$. The goal of this work is to investigate the optimum theoretical description for weak and strong perturbation conditions in near-relativistic collisions, the distinctive features of which are visible in the measured ELC cusp shape.

We briefly mention that the process termed ELC in literature refers to projectile ionization, while the process termed electron capture to continuum (ECC) refers to target ionization with a subsequent capture of the electron into the projectile continuum \cite{rudd_energy_1966,macek_theory_1970,crooks_experimental_1970,dettmann_charge_1974,macek_evidence_1981,breinig_experiments_1982}. While the ECC is not in the focus of the present study, the relationship between both processes will be discussed later in the paper.

The paper is organized as follows: Sec.~\ref{sec:experiment} provides a brief description of the experiment, Sec.~\ref{sec:theory} describes the applied theoretical approach, and in Sec.~\ref{sec:results} we discuss our experimental and theoretical results. 

\section{\label{sec:experiment}Experiment}

The experiment was performed at the heavy-ion accelerator complex of GSI Helmholtzzentrum f\"ur Schwerionenforschung in Darmstadt, Germany. Uranium ions were accelerated by the linear accelerator UNILAC and synchrotron SIS18 to the desired beam energy, and stripped to the Li-like charge state by traversing a  29~mg/cm$^2$ carbon stripper foil. After injection into the experimental storage ring (ESR), electron cooling was applied to the ion beam. The projectile beam energy was defined by the space-charge corrected electron cooler voltage of $41.64$~kV, resulting in the beam energy of 75.91~MeV/u. This corresponds to a projectile velocity in units of speed of light, $v_{\rm p}=0.3808~c$, and a Lorentz factor of $\gamma=1.081$. From the cooler voltage we obtain $E_0=41.64$~keV as the kinetic energy of an electron whose velocity is identical to the projectile velocity. After initial electron cooling, the supersonic gas-jet target was activated at an average area density of $2\times10^{12}$ and $7\times10^{10}~\mathrm{atoms}/\mathrm{cm}^2$ for N and Xe, respectively. The mean number of ions averaged over the measurement phases was $2\times10^7$, while the circulation frequency of the ions was 1.0 MHz. The interaction point was defined by the overlap volume between the ion beam and the gas-jet target. 

Electrons emitted from the interaction point into the forward direction were detected by the electron spectro\-meter within a polar acceptance angle of $\vartheta_\mathrm{e}=0^\circ - \vartheta_{\rm max}=0^\circ - 3.3^\circ$ and the full azimuthal acceptance angle of $\varphi_\mathrm{e}=0^\circ-360^\circ$ with respect to the projectile beam \cite{hillenbrand_radiative_2020}. The electron spectrometer consisted of a sequence of a $60^\circ$ dipole magnet, an iron-free quadrupole triplet, and another $60^\circ$ dipole magnet \cite{hillenbrand_electron-loss--continuum_2014}. Through this highly-selective imaging system a narrow momentum interval of the electron spectrum was guided from the interaction point onto a position sensitive electron detector, more specifically a micro-channel-plate (MCP) detector with a delay-line anode. The flight path from the interaction point to the detector was $4.17$~m in length. The electron momentum spectrum was measured by incrementally scanning the magnetic fields of the spectrometer over the investigated momentum range. In parallel, singly ionized U$^{90+}$ ions were measured by a multi-wire proportional counter (MWPC) positioned after the ring's dipole magnet downstream the target. Electrons attributed to the ELC process were identified as coincidences with the detected U$^{90+}$ ions. For each magnetic setting, the number of coincidence events, $N_{\mathrm{e}\wedge\mathrm{loss}}$, -- corrected for random coincidences -- was normalized to the total number of detected U$^{90+}$ ions, $N_\mathrm{loss}$, which was taken as a measure for the integrated luminosity. By this, the detection efficiency of the particle detector, which is close to unity, cancels out.

The experimentally derived DDCS were evaluated on a relative scale by
\begin{equation}\label{eq:analysis}
	\left. \frac{d^2\sigma}{dE_\mathrm{e} d\Omega_\mathrm{e}}\right|_{\vartheta = 0^\circ} \sim \frac{N_{\mathrm{e}\wedge\mathrm{loss}}}{N_\mathrm{loss}} \frac{1}{\epsilon_\mathrm{e} \Delta\Omega_\mathrm{e}}\frac{E_\mathrm{e}+m_\mathrm{e}c^2}{E_\mathrm{e}^2+2E_\mathrm{e} m_\mathrm{e}c^2} \frac{1}{\Delta p_\mathrm{e}/p_\mathrm{e}}.
\end{equation}
The energy factor with the electron rest energy, $m_\mathrm{e}c^2$, accounts for the transformation of the DDCS from momentum to energy space \cite{hillenbrand_electron-loss--continuum_2014}. The electron detector efficiency, $\epsilon_\mathrm{e}$, the acceptance angle, $\Delta\Omega_\mathrm{e}=\pi\vartheta_\mathrm{max}^2$, and the relative momentum acceptance, $\Delta p_\mathrm{e}/p_\mathrm{e}\approx 0.02$, were assumed to be energy-independent. Possible deviations from this assumption were accounted for by a relative systematic uncertainty of 10\%. The relative total error  includes the systematic and a minor statistical error. 

The magnetic fields applied to the dipole magnets of the electron spectrometer were measured on a relative scale by Hall probes. This momentum scale was converted into an energy scale and subsequently calibrated based on the symmetry of the electron cusp around $E_0$ for reaction (\ref{eq:Ntarget}). In order to verify that the energy calibration was maintained when switching from the measurement of reaction (\ref{eq:Ntarget}) to reaction (\ref{eq:Xetarget}), we analyzed the time of flight of the electrons for both systems. The time of flight was measured as the time difference between the MWPC signal for the ionized projectile and the MCP signal for the electron. The nominal time of flight ranged from 51 to 24~ns for $E_\mathrm{e}$ ranging from 20 to 120~keV, respectively. Both detectors had timing resolutions of a few  ns. In combination with good statistics the time of flight for each data point could be determined with sub-ns precision. However, due to different signal propagation times through the data acquisition, the offset of the time of flight was unknown and had to be deduced from the comparison with the relative Hall-probe measurements. By this complementary method, the energy axis of the electron spectra was determined with an accuracy of  $\delta E_\mathrm{e}/E_\mathrm{e}\approx 2\times\delta p_\mathrm{e}/p_\mathrm{e} <\pm2\%$, and we could confirm that the energy calibration did not change between the settings for both systems, e.g., due to potential hysteresis effects.

During the experiment, the process of radiative electron capture to continuum (RECC) was measured in parallel. Here, a target electron is captured into the projectile continuum, and the excess energy is emitted by a photon,
\begin{equation} \label{eq:recc}
	{\rm U}^{89+} + {\rm N} \rightarrow {\rm U}^{89+}+{\rm N}^+ + {\rm e}^-(E_\mathrm{e},\vartheta_\mathrm{e}) + {\rm \gamma}(E_{\rm \gamma},\vartheta_{\rm \gamma}).
\end{equation}
This process corresponds to the high-energy endpoint of electron-nucleus bremsstrahlung in inverse kinematics \cite{hillenbrand_radiative-electron-capture--continuum_2014}. Reaction (\ref{eq:recc}) was analyzed in Ref.~\cite{hillenbrand_radiative_2020}. The electron spectrum of the RECC has a very steep rise at $E_0$. The good agreement of the experimental and theoretical results for this steep rise provided a solid confirmation for the electron energy calibration. 

Besides the ionized U$^{90+}$ ions, the recombined U$^{88+}$ ions were counted for diagnostic purposes. The cross section for (radiative + nonradiative) electron capture from the target atom into the U$^{89+}$ projectile increases more rapidly with $Z_\mathrm{t}$ than the electron loss cross section. The observed cross section ratio $\sigma_\mathrm{loss}/\sigma_\mathrm{cap}$ was about 0.48 and 0.03 for the N$_2$ and the Xe target, respectively. This illustrates one of the main challenges for the experiment: For the Xe target, most of the projectile ions are lost to the capture channel instead of the investigated ionization channel.  

\section{\label{sec:theory}Theory}

In the considered process the projectile ion initially has three electrons, U$^{89+}(1s^22s)$. In general, each of these electrons can contribute to the ELC emission spectra. However, according to work \cite{bondarev_calculations_2020} the contribution of $K$-shell electrons is less than 4\%. Thus, the participation of the ionic $1s$ electrons in the process can be reduced to a simple screening of the uranium nucleus. 

To describe the $2s$ electron and the emitted electron in the field of the screened ionic nucleus the Furry picture is used, where the corresponding interaction is taken into account nonperturbatively from onset. The interaction between the ionic electrons and the atomic (target) nucleus is considered within two different approximations. The first one is the first-order approximation, in which the interaction between the ionic electron and the atomic nucleus is described within the first order of perturbation theory \cite{voitkiv_plane-wave_2000,lyashchenko_electron_2018}. The second one is the continuum-distorted-wave - eikonal-initial-state (CDW-EIS) approximation, in which this interaction is approximately taken into account in all orders \cite{voitkiv_three-body_2007,voitkiv_electron_2007}. These two approximations are described in Sec.~\ref{sec:theoA} and \ref{sec:theoB}, respectively.

Participation of the atomic electrons in the process is dual. On the one hand they induce a screening of the atomic nucleus field, which reduces the electron loss cross section. This effect is termed elastic target mode. On the other hand, the impact of the atomic electrons can also lead to electron loss from the ion, which increases the corresponding cross section. This effect is termed inelastic target mode. The elastic target mode is discussed below for each approximation separately. 

If the characteristic momenta of all atomic electrons are much smaller than the characteristic momentum transferred to the ion during the collision as well as the characteristic momentum transferred to the atom, then the atom participating in the ELC can be considered as a set of noninteracting particles moving with the same velocity. In this case the following estimate can be done. In such process, the nucleus acts as a single particle with a charge $Z_\mathrm{t}$. If $\nu_{\rm t} \ll 1$, then its contribution to the cross section is proportional to $Z_\mathrm{t}^2$, while the electrons act as $Z_\mathrm{t}$ independent particles and their collective contribution is proportional to their number $Z_\mathrm{t}$. In the case of a collision with a nitrogen atom these conditions are met, and the corresponding estimation for the contribution of the inelastic target mode is 14\%. However, for collision with a xenon atom, these conditions -- strictly speaking -- are not met. Nonetheless we can expect that the effect of the inelastic target mode is even smaller for xenon than for nitrogen. Moreover, since the collision energy under consideration is only slightly higher than the threshold for electron loss by electron impact located at about 60~MeV/u, it is additionally expected that this contribution is small. 

\subsection{First-order approximation}\label{sec:theoA}
The theoretical description of the ELC is conveniently considered in the ion rest frame, where the ion nucleus is placed at the origin. To describe the motion of an atomic nucleus, we use a semiclassical approach, within which the nucleus is treated as a charged particle moving along the $z$-axis with velocity $v_\mathrm{p}$ and has the impact parameter ${\bm b}$.

At first we consider the ELC for a collision with a bare nucleus. The 4-vector potential generated by the moving atomic nucleus can be expressed as 
\begin{equation}
	A^{\mu}(x)
	=\label{eq_A1}
	(A_0, {\bm A})
	\,,
\end{equation}
\begin{equation}
	A_{0}
	=\label{eq_A2}
	0;
	\phantom{12}
	{\bm A} = 
	-\frac{e Z_{\rm t}}{v_{\rm p}s}\left(
	\frac{s_x}{s+s_z},\, \frac{s_y}{s+s_z}, \, \frac{1}{\gamma}
	\right)
	\,,
\end{equation}
where ${\bm{s}}=({\bm{r}}_{\perp}-{\bm{b}}, \, \gamma(z-v_\mathrm{p}t))$ is a distance between the atomic nucleus and the ionic electron with radius vector ${\bm{r}}$, $\gamma=1/\sqrt{1-v_\mathrm{p}^2}$ is the Lorentz factor of the collision, and $e>0$ is the elementary charge. Relativistic units ($\hbar=1$, $c=1$, $m_{\rm e}=1$) are used throughout this section, except otherwise stated. We note that Eq.~(\ref{eq_A2}) represents the Li\'enard-Wiechert potential, 
\begin{equation}
	\tilde A_{0}=
	\frac{eZ_{\rm t}\gamma}{s};
	\phantom{12}
	\tilde{\bm A} = 
	\tilde A_{0}{\bm v}_{\rm p}
	\,,
\end{equation}
with the gauge transformation $A^{\mu}=\tilde A^{\mu}+\partial^{\mu}\chi$ given by the function $\chi= \frac{eZ_{\rm t}}{v_{\rm p}}\ln(s+s_z)$.

In the first order of the perturbation theory with respect to the interaction between the ionic electron and the bare nucleus, the amplitude of the electron loss reads as  
\begin{equation}\label{eq_ampl_1st}
	a^{(1)}({\bm b})=
	-i e \int d^4 x \, {\bar{\psi}}_{\rm f}({\bm r}) \gamma^{\mu}A_{\mu}(x) \psi_{\rm i}({\bm r}) 
	e^{it(\varepsilon_{\rm f}-\varepsilon_{\rm i})}
	\,,
\end{equation}
where the initial and final states of the system are described by one-electron wave functions, $\psi_{\rm i}$ and $\psi_{\rm f}$, respectively, which are eigenvectors of the Dirac equation with the external Coulomb field of the screened ionic nucleus. Furthermore, in Eq.~(\ref{eq_ampl_1st}) $\varepsilon_{\rm i}$ and $\varepsilon_{\rm f}$ denote the initial and the final energy of the ionic electron, respectively, $x^{\mu}=(t, {\bm r})$, and $\gamma^{\mu}$ denotes the Dirac gamma matrices. For the numerical calculations it is convenient to work with the amplitude in momentum space, $S^{(1)}({\bm Q})$, which is connected with the amplitude of Eq.~(\ref{eq_ampl_1st}) by Fourier transformation \cite{voitkiv_three-body_2007},
	\begin{eqnarray}
		&&S^{(1)}({\bm Q})
		=\label{eq_s_1st}
		\frac{1}{2\pi}
		\int d^2 {\bm b} \, e^{i {\bm Q}\cdot{\bm b}}  a^{(1)}({\bm b})
		=
		\frac{2i e^2 Z_{\rm t}}{v_{\rm p}^2} \frac{1}{q'^2 q_z}\\
		&&\times\left(
		\langle \psi_{\rm f} | e^{i{\bm q}\cdot{\bm r}} 
		\left( q_x \alpha_x + q_y \alpha_y \right)
		| \psi_{\rm i}\rangle
		+
		\frac{1}{\gamma^2} \langle \psi_{\rm f}| e^{i{\bm q}\cdot{\bm r}}  q_z \alpha_z | \psi_{\rm i}\rangle
		\right)
		\,,\nonumber
	\end{eqnarray}
where ${\bm{q}}=({\bm Q}, q_z)$ denotes the momentum transfer in the collision, $q_z=\frac{\varepsilon_{\rm f}-\varepsilon_{\rm i}}{v_{\rm p}}$, ${\bm{q}}'=({\bm Q}, q_z/\gamma)$, and $\alpha_x$, $\alpha_y$, and $\alpha_z$ are the Dirac alpha matrices.

To account for the screening effect, i.e., the elastic target mode, we replace the potential of the bare nucleus Eq.~(\ref{eq_A2}) with a sum of Yukawa potentials, which describes the potential of the neutral atom. The expression for the ELC amplitude with this potential has the same form as Eq.~(\ref {eq_s_1st}) with the replacement of the nucleus charge $e Z_{\rm t}$ by the effective charge $e Z_{\rm t}^{\rm eff}({\bm{Q}})$ \cite{voitkiv_relativistic_2008}, where
\begin{equation} \label{eq_z_eff}
	Z_{\rm t}^{\rm eff}({\bm{Q}}) =
	Z_{\rm t} \left( {{Q}^2} +  \frac{q_{z}^2}{\gamma^2}  \right) \sum_{i=1}^3 \frac{A_i}{k_i^2+{{Q}^2} + \frac{q_{z}^2}{\gamma^2}}
	\,\,.
\end{equation}
The parameters $A_i$ and $k_i$ ($i=1,2,3$) are tabulated for various atoms in \cite{moliere1947Naturforsch133,salvat1987pra467}.

The differential cross section of the ELC is connected with the amplitude of Eq.~(\ref{eq_s_1st}) as,
\begin{equation}
	\frac{d^2\sigma}{d E'_{\rm e} d\Omega'_{\rm e}}
	=\label{eq_sigma}
	\frac{E'_{\rm e} p'_{\rm e}}{(2\pi)^3}\int d^2{\bm Q}\, \left| S^{(1)}({\bm Q})\right|^2,
\end{equation}
where $E_{\rm e}'\equiv\varepsilon_{\rm f}$ is the energy of the emitted electron in the ion rest frame, $p'_{\rm e}$ is the corresponding momentum, and $\Omega'_{\rm e}$ is the solid angle defining the direction of the emission. To compare the theoretical and experimental results, we transfer the differential cross section of Eq.~(\ref{eq_sigma}) to the atom rest frame and average over the angular acceptance of the spectrometer
\begin{equation}\label{eq:cusp}
	\left. \frac{d^2 \sigma}{dE_\mathrm{e} d\Omega_\mathrm{e}} \right|_{\vartheta_\mathrm{e}=0^\circ} = \frac{1}{1-\cos \vartheta_\mathrm{max}} \int_0^{\vartheta_\mathrm{max}} \frac{d^2 \sigma}{dE_\mathrm{e}' d\Omega_\mathrm{e}'} \sin \vartheta_\mathrm{e}'~d\vartheta_\mathrm{e}.
\end{equation}
The transformation includes a rotation of the polar angle
by 180$^{\circ}$,
\begin{equation}
	\vartheta_\mathrm{e}'(E_\mathrm{e},\vartheta_\mathrm{e})=180^\circ-\arctan\left[ \frac{\sin \vartheta_\mathrm{e}}{\gamma [\cos \vartheta_\mathrm{e} - v_\mathrm{p}/v_\mathrm{e}]} \right].
\end{equation}
Therefore, forward emitted electrons in the projectile frame, i.e., $\vartheta'_{\rm e}<90^\circ$, correspond to electron energies $E_{\rm e}<E_0$ and electron velocities $v_{\rm e}<v_{\rm p}$ in the target frame, and vice versa.

Note, that the trivial (i.e., purely kinematic) cusp shape
is given by
\begin{equation}\label{eq:0}
	\left. \frac{d^2 \sigma^\mathrm{kin.}}{dE_\mathrm{e} d\Omega_\mathrm{e}} \right|_{\vartheta_\mathrm{e}=0^\circ} \sim \frac{1}{1-\cos \vartheta_\mathrm{max}} \int_0^{\vartheta_\mathrm{max}} \sin \vartheta_\mathrm{e}'~d\vartheta_\mathrm{e}.
\end{equation}

\begin{figure*}[!t]
	\subfigure{\includegraphics[width=0.66\columnwidth]{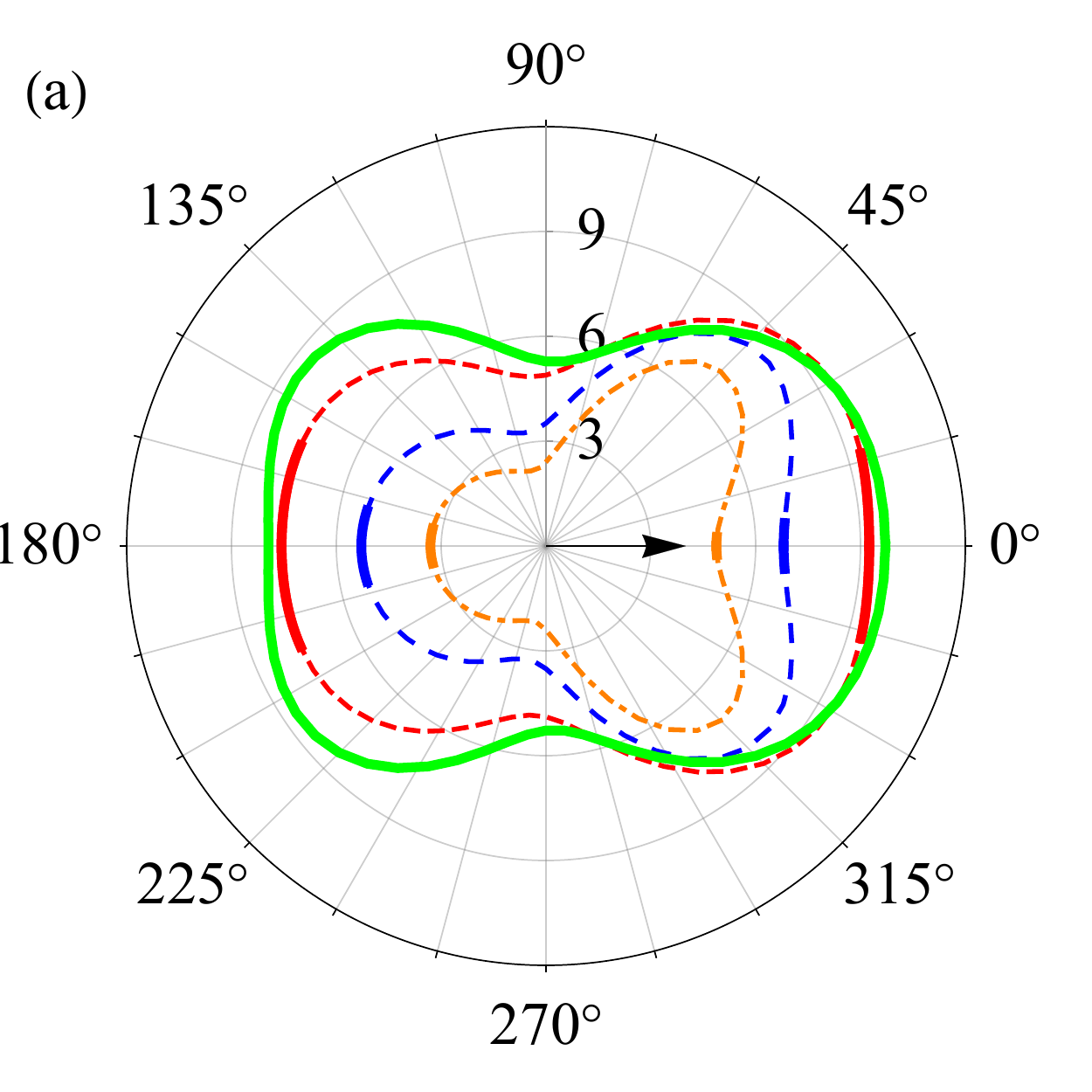}}	
	\subfigure{\includegraphics[width=0.66\columnwidth]{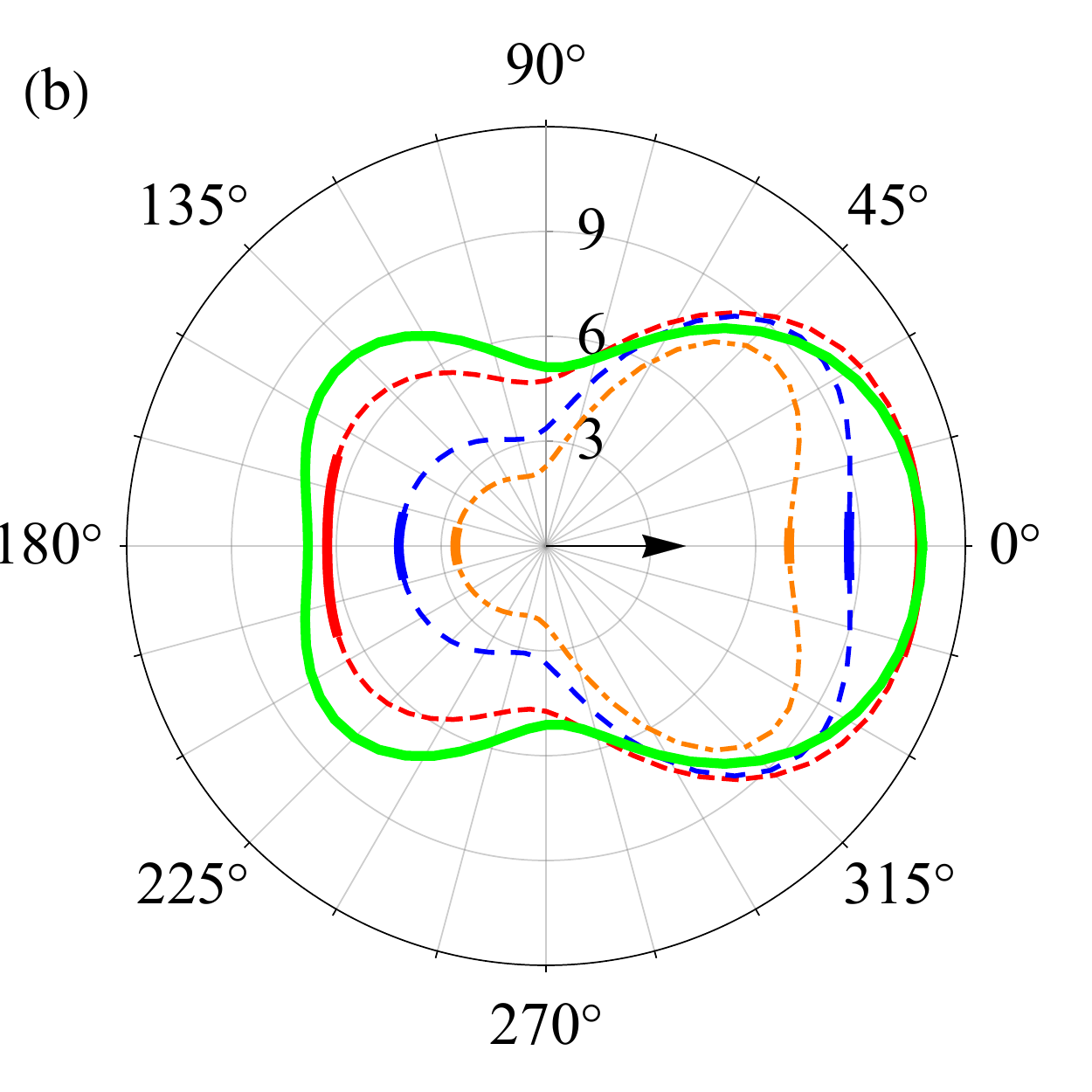}}	
	\subfigure{\includegraphics[width=0.66\columnwidth]{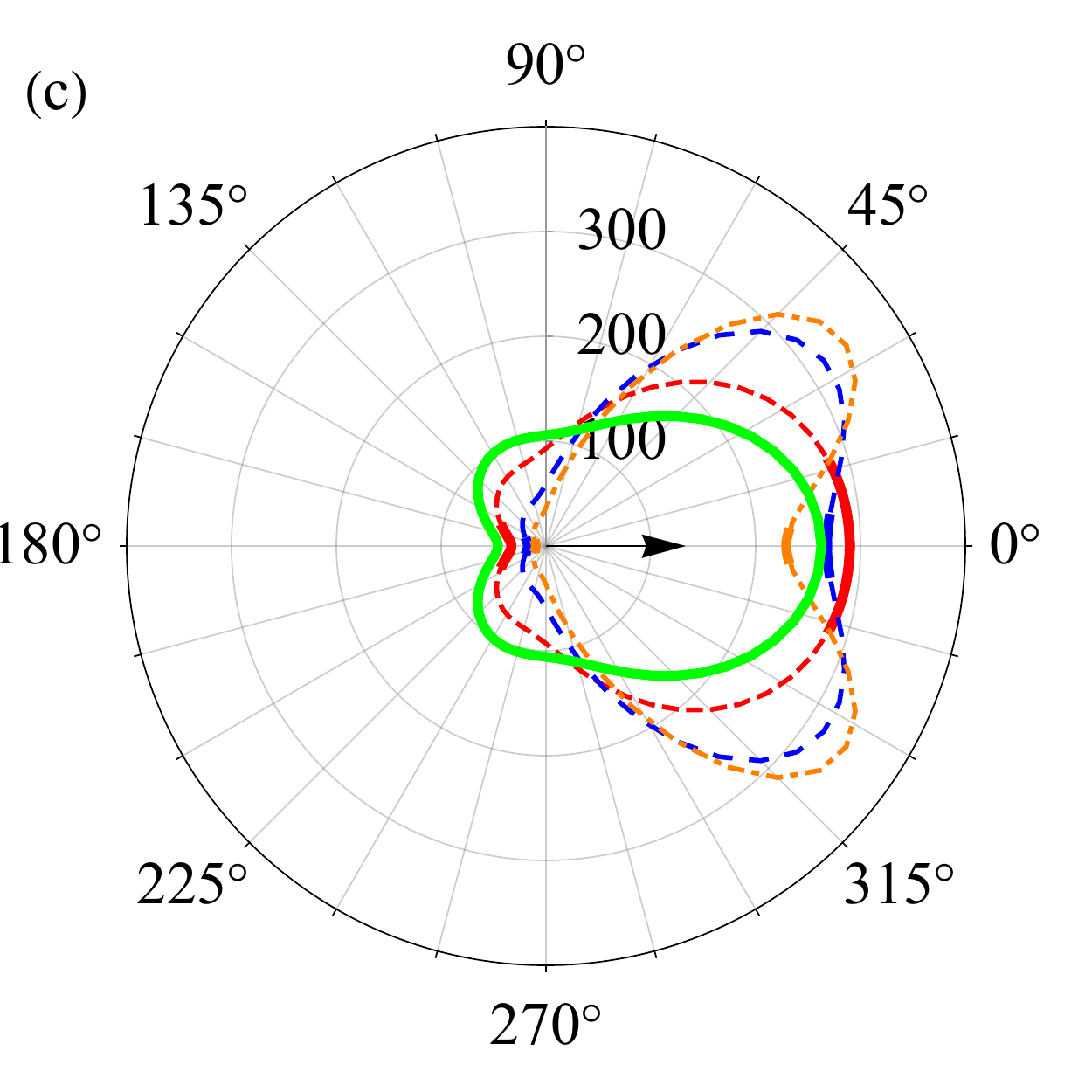}}	
	\caption{\label{fig:theo} DDCS (barn/keV/sr) in the U$^{89+}$-projectile frame as a function of the energy, $E_\mathrm{e}'$, and the polar angle, $\vartheta_\mathrm{e}'$, of the emitted electron: (a)~first-order approximation for a N$^{7+}$ target, (b)~CDW-EIS approximation for a N$^{7+}$ target, (c)~CDW-EIS approximation for a Xe$^{54+}$ target. The energy, $E_\mathrm{e}'$, was chosen to be 0.1~keV (solid green line), 1~keV (short-dashed red line), 5~keV (long-dashed blue line), and 10~keV (dot-dashed orange line). The direction of motion of the ionizing nucleus is indicated with the arrow. The polar angles that fall into the acceptance of the spectrometer are marked in solid bold lines. At small electron emission energies, such as $E_\mathrm{e}'=0.1$~keV, all values of $\vartheta_\mathrm{e}'$ fall into the acceptance of the spectrometer.}
\end{figure*}

If the expansion parameter of the perturbation theory is much smaller than 1 ($\nu_{\rm t}=\alpha Z_{\rm t}/v_{\rm p}\ll 1$), then the first-order approximation provides an accurate description of the process. For $v_{\rm p}=0.3808$~r.u. and a nitrogen target the expansion parameter is $\nu_{\rm t}=0.13$, which justifies the use of this approximation. However, in the case of the xenon target and the same collision velocity, the expansion parameter is $\nu_{\rm t}=1.04$, which exceeds the limitation for this approximation. For $\nu_{\rm t}\approx 1$ the interaction between the ionic electron and the atomic nucleus should be considered nonperturbatively. For this purpose the CDW-EIS approximation is employed.

\subsection{CDW-EIS approximation}\label{sec:theoB}
Within the CDW-EIS approximation, the initial (bound) and final (emitted) electrons are described by the following functions \cite{voitkiv_three-body_2007},
\begin{eqnarray}\label{eq_chi_i}
	\chi_{\rm i}(x)
	&=&
	\psi_{\rm i}(\bm r)e^{-i\varepsilon_{\rm i} t} (v_{\rm p}s+ {\bm v}_{\rm p}\cdot{\bm s})^{-i\nu_{\rm t}}
	\,,\\
	\chi_{\rm f}(x)
	&=&
	\psi_{\rm f}(\bm r)e^{-i\varepsilon_{\rm f} t}\label{eq_chi_f} \\
	&\times&
	\Gamma(1+i\eta_{\rm t}) e^{\pi \eta_{\rm t}/2} 
	F(-i\eta_{\rm t},1,-i(p_{\rm e}s+ {\bm p}_{\rm e}\cdot{\bm s}))
	\,,\nonumber
\end{eqnarray}
where $\eta_{\rm t}=\alpha Z_{\rm t}/v_{\rm e}$, $v_{\rm e}$ and ${\bm p}_{\rm e}$ are the velocity and momentum of the emitted electron in the rest frame of atomic nucleus, respectively. The functions $\Gamma(z)$ and $F(a,b,z)$ in Eq.~(\ref{eq_chi_f}) denote the Gamma function and the confluent hypergeometric function, respectively.

\begin{figure*}[t!]
	\subfigure{\includegraphics[width=1\columnwidth]{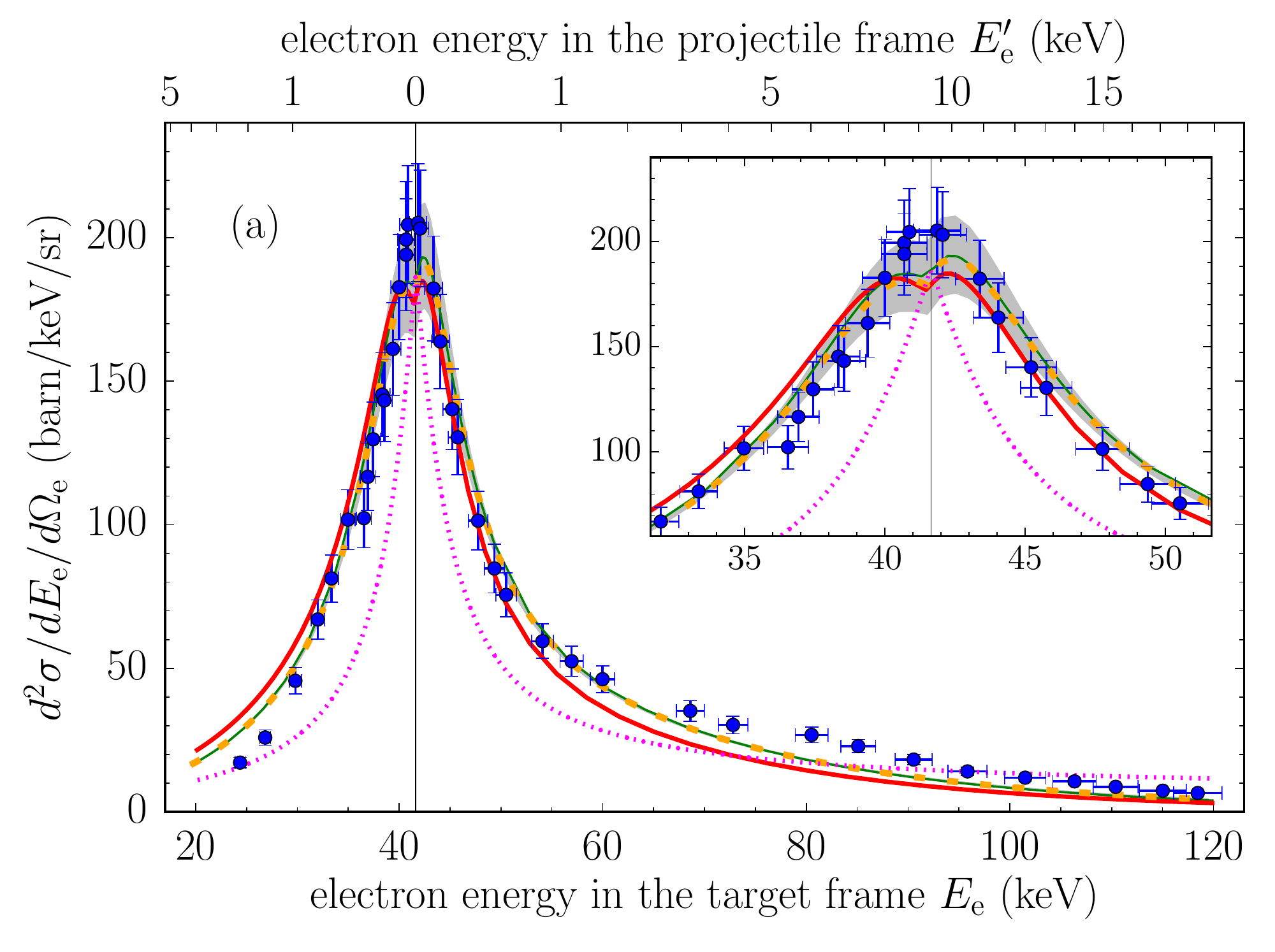}}	
	\subfigure{\includegraphics[width=1\columnwidth]{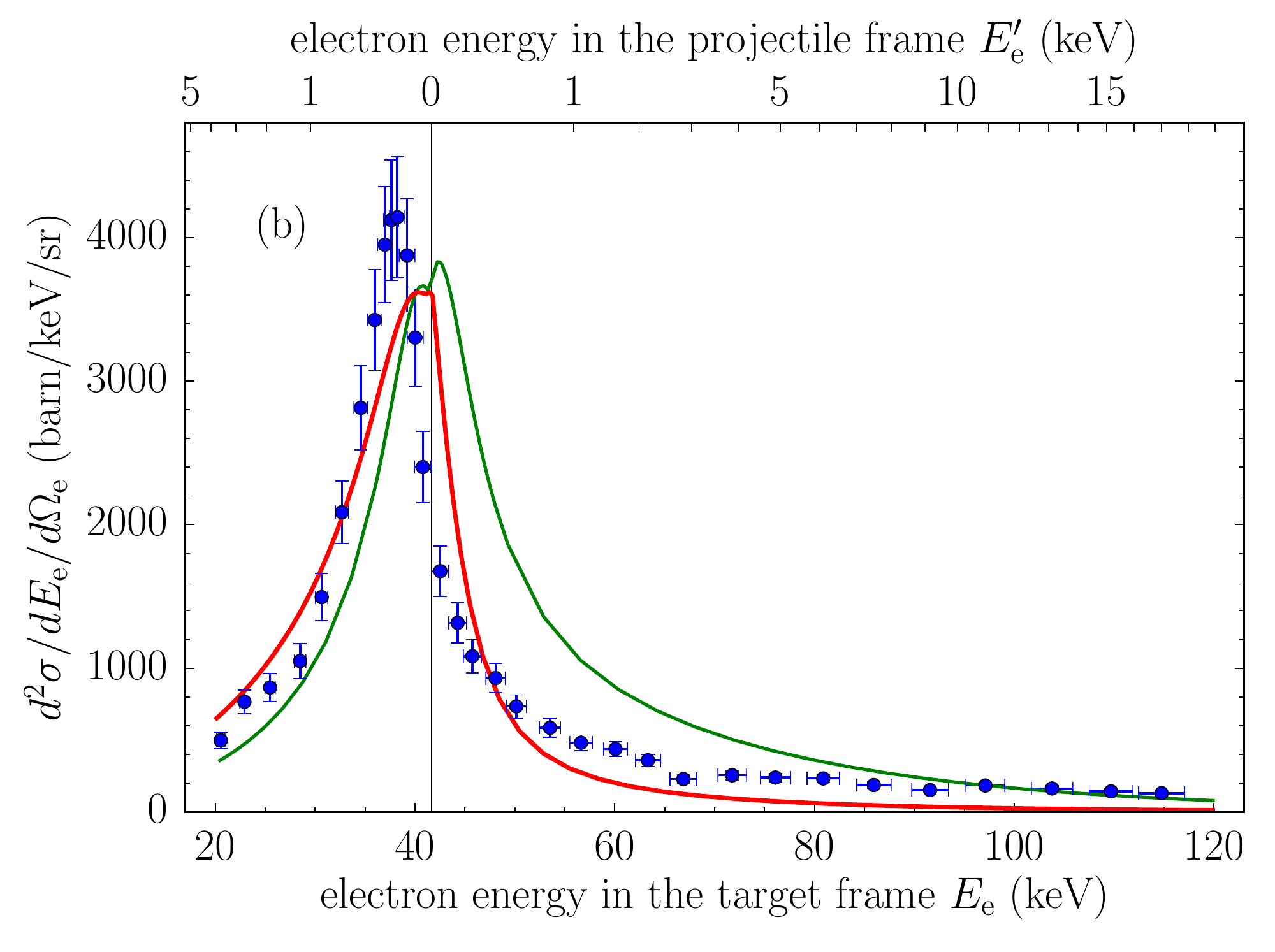}}	
	\caption{\label{fig:exp} DDCS in the target frame for (a) ${\rm U}^{89+} + {\rm N}$ and (b)  ${\rm U}^{89+} + {\rm Xe}$: Experimental data points (blue dots with systematic error bars) were normalized to theoretical CDW-EIS results (thick red line). The present first-order results shown in the thin green line are in excellent agreement with the theory of Ref.~\cite{bondarev_calculations_2020} shown as the dashed orange line. The latter study includes a variation of the acceptance angle of the spectrometer, $\vartheta_{\rm max}=3.3\pm0.3^\circ$, shown as the gray area. In (a) the dotted magenta line shows the purely kinematic cusp shape according to Eq.~(\ref{eq:0}), normalized to theory at $E_0$. In (b) the theoretical first-order DDCS was divided by a factor 3 in order to match the range of the plot.}
\end{figure*}	

Replacing the Dirac wave functions in Eq.~(\ref{eq_ampl_1st}) by functions (\ref{eq_chi_i}) and (\ref{eq_chi_f}) we obtain the following expressions for the amplitudes in coordinate and momentum spaces within the CDW-EIS approximation: 
\begin{widetext}
	\begin{eqnarray}
		a^{(\rm{CDW})}({\bm{b}})
		&=&\nonumber
		\frac{i e^2 Z_{\rm t}}{v_{\rm p}}
		\int dt \, e^{it(\varepsilon_{\rm f}-\varepsilon_{\rm i})}
		\int d^3 {\bm{r}} \, \Gamma(1-i\eta_{\rm t}) e^{\pi\eta_{\rm t}/2} F(i\eta_{\rm t},1,i(p_{\rm e}s+{\bm{p}_{\rm e}}\cdot{\bm{s}})) (v_{\rm p}s+{\bm{v}}_{\rm p}\cdot{\bm{s}})^{-i\nu_{\rm t}}\\
		&\times&
		\psi_{\rm f}^{+}({\bm{r}})
		\frac{1}{s}\left(
		\frac{s_x\alpha_x+s_y\alpha_y}{s+s_z} + \frac{\alpha_z}{\gamma}
		\right)
		\psi_{\rm i}({\bm{r}})
		\,,
	\end{eqnarray}
	\begin{eqnarray}
		S^{(\rm{CDW})}({\bm{Q}})
		&=&\label{eq_s_cdw}
		\frac{2i e^2 Z_{\rm t} }{v_{\rm p}\gamma} 
		\left( i\frac{A-B}{A}\right)^{-i\eta_{\rm t}}
		\left( i\frac{A}{C}\right)^{i\nu_{\rm t}}
		\frac{\Gamma(1-i\eta_{\rm t})\Gamma(1-i\nu_{\rm t})}{A}
		\langle\psi_{\rm f}|
		e^{i{\bm{q}}\cdot{\bm{r}}} 
		\left(J_x\alpha_x + J_y\alpha_y + J_z\alpha_z\right) | \psi_{\rm i}\rangle
		\,,
	\end{eqnarray}
	where
	\begin{eqnarray}
		J_{x(y)}
		&=&
		\frac{2}{ C}
		\left[
		q_{x(y)} {_2}F_1\left(i\eta_{\rm t},i\nu_{\rm t},1, \frac{BC-AD}{C(B-A)} \right) 
		+i\eta_{\rm t} \frac{A(-Cp_{{\rm e},x(y)} + Dq_{x(y)})}{C(A-B)}{_2}F_1\left(i\eta_{\rm t}+1,i\nu_{\rm t}+1,2, \frac{BC-AD}{C(B-A)} \right) 
		\right]
		\,,
	\end{eqnarray}
\end{widetext}

\begin{equation}
	J_{z}=
	\frac{1}{\gamma v_{\rm p}}
	{_2}F_1\left(i\eta_{\rm t}, i\nu_{\rm t}, 1, \frac{BC-AD}{C(B-A)}\right)
	\,,
\end{equation}
\begin{equation}
	A=
	q'^2
	\,, \,
	B=2\bm{q}'\cdot\bm{p}_{\rm e}
	\,, \,
	C=2q'_zv_{\rm p}
	\,, \,
	D=2v_{\rm p}(p_{{\rm e},z}-p_{\rm e})
	\,,
\end{equation}
and function $_2F_1(a,b,c,z)$ denotes the hypergeometric function.
We note that Eq.~(\ref{eq_s_cdw}) reduces to Eq.~(\ref{eq_s_1st}) for $\nu_{\rm t}\to 0$.

The CDW-EIS approximation is formulated in such a way [see Eqs.~(\ref{eq_chi_i} and (\ref{eq_chi_f})] that the screening of the atomic nucleus by its electrons cannot be taken into account by the simple method used in the first-order approximation [see Eq.~(\ref{eq_z_eff})]. In the present study, we do not take into account the screening effect within the CDW-EIS approximation. However, due to the large momentum transfers involved in the electron loss process (corresponding to quite small impact parameters) the screening effects are expected to be very small. (In particular, according to our first-order calculation, even for the Xe target, they do not exceeding a few percent.) Moreover, in the case of electron loss from very heavy ions the CDW-EIS approximation using the Dirac wave functions represents a significant improvement \cite{voitkiv_three-body_2007} compared to approaches that applied semirelativistic Darwin and Sommerfeld-Maue-Furry approximations,
 e.g.~\cite{deco_asymptotic_1989,deco_k-shell_1990}.

\section{Results}\label{sec:results}

Our calculated spectra for the theoretical DDCS in the projectile frame are shown in Fig.~\ref{fig:theo}. For the ionization of the U$^{89+}$ ion by a N$^{7+}$ nucleus, the first-order results presented in Fig.~\ref{fig:theo}(a) and the CDW-EIS results presented in Fig.~\ref{fig:theo}(b) are rather similar. This is consistent with the fact that the CDW-EIS results converge towards the first-order results in the limit of weak perturbations  $\nu_{\rm t}\to 0$ \cite{voitkiv_electron_2007}. The first-order results for the ionization by Xe$^{54+}$ are not shown, because they are just the DDCS of the nitrogen target [presented in Fig.~\ref{fig:theo}(a)] multiplied by the factor $54^2/7^2 \approx 60$. The CDW-EIS results for Xe$^{54+}$ shown in Fig.~\ref{fig:theo}(c) exhibit a strong asymmetry along the collision axis with a pronounced preference for forward emission angles. This behavior illustrates how the electrons ionized from the U$^{89+}$ ion are attracted by the transient Xe$^{54+}$ nucleus, therefore it is a clear signature of a two-center effect. We note that in the case of collision with Xe$^{54+}$ the first-order approximation considerably overestimates the differential and total cross section with respect to the result of the CDW-EIS approximation. This behavior was analyzed for total cross sections of comparable collision systems in \cite{voitkiv_three-body_2007,voitkiv_electron_2007}. We also investigated the magnitude of the screening effect within the first-order approximation, where it turned out to be insignificant.

The experimental and theoretical results for the ELC cusp spectra are shown in Fig.~\ref{fig:exp}. The results within the first-order approximation as well as within the CDW-EIS approach are presented. The experimental data were normalized to the CDW-EIS results for both systems individually, while the energy calibration was the same for both systems. As shown in Fig.~\ref{fig:exp}(a), our first-order approximation is in excellent agreement with the results presented in Ref.~\cite{bondarev_calculations_2020}, which were obtained independently. As previously seen in the projectile frame spectra, the first-order and CDW-EIS results are very similar for the weak perturbation of the nitrogen target. For the nitrogen target, the expected symmetry of the electron cusp around $E_0$ is seen in both, the experimental and the theoretical, data. Strictly speaking, this symmetry applies to the electron spectrum in momentum space, while the transformation into energy space [cf.~Eq.~(\ref{eq:analysis})] already introduces a slight asymmetry. This can as well be seen in the trivial cusp shape of Eq.~(\ref{eq:0}) shown in Fig.~\ref{fig:exp}(a). All experimental and theoretical results clearly differ from the trivial cusp shape. 

For our results with the xenon target shown in Fig.~\ref{fig:exp}(b), the centroid of the electron cusp distribution is shifted distinctly below $E_0$ in both, the experimental data and the CDW-EIS results, though to different values. For a better comparison of the cusp shape, the first-order DDCS are divided by an arbitrary factor three, which leads to a coincidental agreement at large $E_\mathrm{e}$. Overall, the CDW-EIS results show a qualitatively better agreement with the experimental data than the first-order results for the strong perturbation caused by the xenon target.

The remaining discrepancy between the experiment and the CDW-EIS prediction may be due to a combination of the following reasons. First,
the distortion of the electron motion due to the field of the atom is taken in a simplified form, corresponding to neglecting the term with the gradient in the Furry wave function~\cite{voitkiv_three-body_2007}. Second,
the present form of the CDW-EIS approximation was shown in Ref.~\cite{voitkiv_three-body_2007} to work quite well for obtaining the total cross section when the perturbation parameter $\nu_{\rm t} \approx 1$ (or even somewhat exceeds unity), however, it might no longer be the case when differential cross sections are considered. Third, we could not exclude that the screening effect, which -- according to the first-order calculation -- is very weak, being properly included into the CDW-EIS approximation might affect the differential cross sections much more than in the first-order one.

The discussion of the presently studied ELC cusp originating from projectile ionization necessitates a reference to the electron-capture-to-continuum (ECC) cusp originating from target ionization \cite{hillenbrand_electron-capture--continuum_2015},
\begin{equation}
	{\rm U}^{89+} + {\rm N} \rightarrow {\rm U}^{89+}+{\rm N}^+ + {\rm e}^-(E_\mathrm{e}\approx E_0,\vartheta_\mathrm{e}=0^\circ).
\end{equation}
The ECC is characterized by a large momentum transfer to the target nucleus and an asymmetric cusp shape with a centroid shifted also towards lower electron energies \cite{hillenbrand_electron-capture--continuum_2015,shah_shifts_2003,bhattacharya_shifting_2005}. However, due to the coincidence condition in the evaluation of our experimental electron spectra through Eq.~(\ref{eq:analysis}), the results shown in Figs.~\ref{fig:exp}(a) and \ref{fig:exp}(b) do not comprise electrons originating from ECC. We also emphasize, that the asymmetry observed presently for the ELC with the xenon target is opposite to the asymmetry observed for the RECC process of reaction (\ref{eq:recc}) measured simultaneously \cite{hillenbrand_radiative_2020}. During the RECC process the excess energy is released by emitting a bremsstrahlung photon, therefore no momentum transfer to the target nucleus is required.

In this context it is also important to recall that the asymmetry of the ELC cusp originating from ionizing a U$^{28+}$ projectile was observed with a strong preference for the high-energy side of the cusp \cite{hillenbrand_strong_2016}, in contrast to the results of the present work. We are not aware of any other system where the ELC cusp asymmetry exhibits such a strong dependence on the electron configuration of the projectile under study. Since there is presently no theoretical explanation available for the observations made for the complex U$^{28+}$ collision systems, the U$^{89+}$ collision systems studied in this work provide an important intermediate step towards understanding ELC at large perturbations.

\section{Summary}\label{sec:summary}

In summary, we present a detailed experimental and theoretical study for the change of the electron cusp shape and the shift of its centroid energy due to the two-center effects in the ELC for near-relativistic heavy-ion collision systems. Our measurement provides a stringent test for the theoretical description of projectile ionization beyond first-order perturbation. Relativistic CDW-EIS calculations are able to qualitatively describe the observed shift of the cusp centroid towards lower electron energies. The remaining discrepancy may motivate the development of even more sophisticated theoretical approaches of such collision systems. Our experimental and theoretical methodology can also be applied to investigate DDCS for electron-impact ionization of heavy highly-charged ions in inverse kinematics \cite{lyashchenko_electron_2018}.

\begin{acknowledgments}
We thank M.~Steck, S.~Litvinov, and R.~Heß for operating the ESR. This research has been conducted in the framework of the SPARC collaboration. The work of K.N.L. is supported by Chinese Postdoctoral Science Foundation No.~2020M673538. E.P.B., E.D.F., and E.V.P. gratefully acknowledge support by the 'Transnational Access to GSI' (TNA) activity. A.I.B. acknowledges support by the Russian Science Foundation, grant No.~20-62-46006. C.B. acknowledges support by the Bundesministerium für Bildung und Forschung (BMBF) Contract No.~05P15RGFAA. M.V. acknowledges support by BMBF Contract No.~05P15SJFAA.  This work was supported by the Helmholtz-CAS Joint Research Group HCJRG-108 and the European Research Council (ERC) under the European Union’s Horizon 2020, Grant No.~682841 “ASTRUm”.
\end{acknowledgments}

\bibliography{paper2021}

\begin{thebibliography}{47}%
\makeatletter
\providecommand \@ifxundefined [1]{%
 \@ifx{#1\undefined}
}%
\providecommand \@ifnum [1]{%
 \ifnum #1\expandafter \@firstoftwo
 \else \expandafter \@secondoftwo
 \fi
}%
\providecommand \@ifx [1]{%
 \ifx #1\expandafter \@firstoftwo
 \else \expandafter \@secondoftwo
 \fi
}%
\providecommand \natexlab [1]{#1}%
\providecommand \enquote  [1]{``#1''}%
\providecommand \bibnamefont  [1]{#1}%
\providecommand \bibfnamefont [1]{#1}%
\providecommand \citenamefont [1]{#1}%
\providecommand \href@noop [0]{\@secondoftwo}%
\providecommand \href [0]{\begingroup \@sanitize@url \@href}%
\providecommand \@href[1]{\@@startlink{#1}\@@href}%
\providecommand \@@href[1]{\endgroup#1\@@endlink}%
\providecommand \@sanitize@url [0]{\catcode `\\12\catcode `\$12\catcode
  `\&12\catcode `\#12\catcode `\^12\catcode `\_12\catcode `\%12\relax}%
\providecommand \@@startlink[1]{}%
\providecommand \@@endlink[0]{}%
\providecommand \url  [0]{\begingroup\@sanitize@url \@url }%
\providecommand \@url [1]{\endgroup\@href {#1}{\urlprefix }}%
\providecommand \urlprefix  [0]{URL }%
\providecommand \Eprint [0]{\href }%
\providecommand \doibase [0]{https://doi.org/}%
\providecommand \selectlanguage [0]{\@gobble}%
\providecommand \bibinfo  [0]{\@secondoftwo}%
\providecommand \bibfield  [0]{\@secondoftwo}%
\providecommand \translation [1]{[#1]}%
\providecommand \BibitemOpen [0]{}%
\providecommand \bibitemStop [0]{}%
\providecommand \bibitemNoStop [0]{.\EOS\space}%
\providecommand \EOS [0]{\spacefactor3000\relax}%
\providecommand \BibitemShut  [1]{\csname bibitem#1\endcsname}%
\let\auto@bib@innerbib\@empty
\bibitem [{\citenamefont {Stolterfoht}\ \emph {et~al.}(1974)\citenamefont
  {Stolterfoht}, \citenamefont {Schneider}, \citenamefont {Burch},
  \citenamefont {Wieman},\ and\ \citenamefont
  {Risley}}]{stolterfoht_mechanisms_1974}%
  \BibitemOpen
  \bibfield  {author} {\bibinfo {author} {\bibfnamefont {N.}~\bibnamefont
  {Stolterfoht}}, \bibinfo {author} {\bibfnamefont {D.}~\bibnamefont
  {Schneider}}, \bibinfo {author} {\bibfnamefont {D.}~\bibnamefont {Burch}},
  \bibinfo {author} {\bibfnamefont {H.}~\bibnamefont {Wieman}},\ and\ \bibinfo
  {author} {\bibfnamefont {J.}~\bibnamefont {Risley}},\ }\bibfield  {title}
  {\bibinfo {title} {Mechanisms for {Electron} {Production} in 30-{MeV} {O$^{n+}$}
  +{O$_2$} {Collisions}},\ }\href {https://doi.org/10.1103/PhysRevLett.33.59}
  {\bibfield  {journal} {\bibinfo  {journal} {Phys. Rev. Lett.}\ }\textbf
  {\bibinfo {volume} {33}},\ \bibinfo {pages} {59} (\bibinfo {year}
  {1974})}\BibitemShut {NoStop}%
\bibitem [{\citenamefont {Stolterfoht}(1997)}]{stolterfoht_electron_1997}%
  \BibitemOpen
  \bibfield  {author} {\bibinfo {author} {\bibfnamefont {N.}~\bibnamefont
  {Stolterfoht}},\ }\href
  {http://link.springer.com/book/10.1007%2F978-3-662-03480-4} {\emph {\bibinfo
  {title} {Electron emission in heavy ion-atom collisions}}},\ \bibinfo
  {series} {Springer {Series} on {Atomic}, {Optical}, and {Plasma} {Physics}}\
  No.~\bibinfo {number} {20}\ (\bibinfo  {publisher} {Springer},\ \bibinfo
  {address} {Berlin},\ \bibinfo {year} {1997})\BibitemShut {NoStop}%
\bibitem [{\citenamefont {Rudd}\ \emph {et~al.}(1966)\citenamefont {Rudd},
  \citenamefont {Sautter},\ and\ \citenamefont {Bailey}}]{rudd_energy_1966}%
  \BibitemOpen
  \bibfield  {author} {\bibinfo {author} {\bibfnamefont {M.~E.}\ \bibnamefont
  {Rudd}}, \bibinfo {author} {\bibfnamefont {C.~A.}\ \bibnamefont {Sautter}},\
  and\ \bibinfo {author} {\bibfnamefont {C.~L.}\ \bibnamefont {Bailey}},\
  }\bibfield  {title} {\bibinfo {title} {Energy and {Angular} {Distributions}
  of {Electrons} {Ejected} from {Hydrogen} and {Helium} by 100- to 300-{keV}
  {Protons}},\ }\href {https://doi.org/10.1103/PhysRev.151.20} {\bibfield
  {journal} {\bibinfo  {journal} {Phys. Rev.}\ }\textbf {\bibinfo {volume}
  {151}},\ \bibinfo {pages} {20} (\bibinfo {year} {1966})}\BibitemShut
  {NoStop}%
\bibitem [{\citenamefont {Macek}(1970)}]{macek_theory_1970}%
  \BibitemOpen
  \bibfield  {author} {\bibinfo {author} {\bibfnamefont {J.}~\bibnamefont
  {Macek}},\ }\bibfield  {title} {\bibinfo {title} {Theory of the {Forward}
  {Peak} in the {Angular} {Distribution} of {Electrons} {Ejecteed} by {Fast}
  {Protons}},\ }\href {https://doi.org/10.1103/PhysRevA.1.235} {\bibfield
  {journal} {\bibinfo  {journal} {Phys. Rev. A}\ }\textbf {\bibinfo {volume}
  {1}},\ \bibinfo {pages} {235} (\bibinfo {year} {1970})}\BibitemShut {NoStop}%
\bibitem [{\citenamefont {Crooks}\ and\ \citenamefont
  {Rudd}(1970)}]{crooks_experimental_1970}%
  \BibitemOpen
  \bibfield  {author} {\bibinfo {author} {\bibfnamefont {G.}~\bibnamefont
  {Crooks}}\ and\ \bibinfo {author} {\bibfnamefont {M.}~\bibnamefont {Rudd}},\
  }\bibfield  {title} {\bibinfo {title} {Experimental {Evidence} for the
  {Mechanism} of {Charge} {Transfer} into {Continuum} {States}},\ }\href
  {https://doi.org/10.1103/PhysRevLett.25.1599} {\bibfield  {journal} {\bibinfo
   {journal} {Phys. Rev. Lett.}\ }\textbf {\bibinfo {volume} {25}},\ \bibinfo
  {pages} {1599} (\bibinfo {year} {1970})}\BibitemShut {NoStop}%
\bibitem [{\citenamefont {Dettmann}\ \emph {et~al.}(1974)\citenamefont
  {Dettmann}, \citenamefont {Harrison},\ and\ \citenamefont
  {Lucas}}]{dettmann_charge_1974}%
  \BibitemOpen
  \bibfield  {author} {\bibinfo {author} {\bibfnamefont {K.}~\bibnamefont
  {Dettmann}}, \bibinfo {author} {\bibfnamefont {K.~G.}\ \bibnamefont
  {Harrison}},\ and\ \bibinfo {author} {\bibfnamefont {M.~W.}\ \bibnamefont
  {Lucas}},\ }\bibfield  {title} {\bibinfo {title} {Charge exchange to the
  continuum for light ions in solids},\ }\href
  {https://doi.org/10.1088/0022-3700/7/2/012} {\bibfield  {journal} {\bibinfo
  {journal} {J. Phys. B}\ }\textbf {\bibinfo {volume} {7}},\ \bibinfo {pages}
  {269} (\bibinfo {year} {1974})}\BibitemShut {NoStop}%
\bibitem [{\citenamefont {Macek}\ \emph {et~al.}(1981)\citenamefont {Macek},
  \citenamefont {Potter}, \citenamefont {Duncan}, \citenamefont {Menendez},
  \citenamefont {Lucas},\ and\ \citenamefont
  {Steckelmacher}}]{macek_evidence_1981}%
  \BibitemOpen
  \bibfield  {author} {\bibinfo {author} {\bibfnamefont {J.}~\bibnamefont
  {Macek}}, \bibinfo {author} {\bibfnamefont {J.~E.}\ \bibnamefont {Potter}},
  \bibinfo {author} {\bibfnamefont {M.~M.}\ \bibnamefont {Duncan}}, \bibinfo
  {author} {\bibfnamefont {M.~G.}\ \bibnamefont {Menendez}}, \bibinfo {author}
  {\bibfnamefont {M.~W.}\ \bibnamefont {Lucas}},\ and\ \bibinfo {author}
  {\bibfnamefont {W.}~\bibnamefont {Steckelmacher}},\ }\bibfield  {title}
  {\bibinfo {title} {Evidence for {Second}-{Born}-{Approximation}
  {Contributions} to {Continuum}-{Electron} {Capture} by {Positive} {Ions}},\
  }\href {https://doi.org/10.1103/PhysRevLett.46.1571} {\bibfield  {journal}
  {\bibinfo  {journal} {Phys. Rev. Lett.}\ }\textbf {\bibinfo {volume} {46}},\
  \bibinfo {pages} {1571} (\bibinfo {year} {1981})}\BibitemShut {NoStop}%
\bibitem [{\citenamefont {Breinig}\ \emph {et~al.}(1982)\citenamefont
  {Breinig}, \citenamefont {Elston}, \citenamefont {Huldt}, \citenamefont
  {Liljeby}, \citenamefont {Vane}, \citenamefont {Berry}, \citenamefont
  {Glass}, \citenamefont {Schauer}, \citenamefont {Sellin}, \citenamefont
  {Alton}, \citenamefont {Datz}, \citenamefont {Overbury}, \citenamefont
  {Laubert},\ and\ \citenamefont {Suter}}]{breinig_experiments_1982}%
  \BibitemOpen
  \bibfield  {author} {\bibinfo {author} {\bibfnamefont {M.}~\bibnamefont
  {Breinig}}, \bibinfo {author} {\bibfnamefont {S.~B.}\ \bibnamefont {Elston}},
  \bibinfo {author} {\bibfnamefont {S.}~\bibnamefont {Huldt}}, \bibinfo
  {author} {\bibfnamefont {L.}~\bibnamefont {Liljeby}}, \bibinfo {author}
  {\bibfnamefont {C.~R.}\ \bibnamefont {Vane}}, \bibinfo {author}
  {\bibfnamefont {S.~D.}\ \bibnamefont {Berry}}, \bibinfo {author}
  {\bibfnamefont {G.~A.}\ \bibnamefont {Glass}}, \bibinfo {author}
  {\bibfnamefont {M.}~\bibnamefont {Schauer}}, \bibinfo {author} {\bibfnamefont
  {I.~A.}\ \bibnamefont {Sellin}}, \bibinfo {author} {\bibfnamefont {G.~D.}\
  \bibnamefont {Alton}}, \emph{et al.},\ }\bibfield  {title}
  {\bibinfo {title} {Experiments concerning electron capture and loss to the
  continuum and convoy electron production by highly ionized projectiles in the
  0.7 - 8.5-{MeV}/u range transversing the rare gases, polycrystalline solids,
  and axial channels in gold},\ }\href
  {https://doi.org/10.1103/PhysRevA.25.3015} {\bibfield  {journal} {\bibinfo
  {journal} {Phys. Rev. A}\ }\textbf {\bibinfo {volume} {25}},\ \bibinfo
  {pages} {3015} (\bibinfo {year} {1982})}\BibitemShut {NoStop}%
\bibitem [{\citenamefont {Drepper}\ and\ \citenamefont
  {Briggs}(1976)}]{drepper_doubly_1976}%
  \BibitemOpen
  \bibfield  {author} {\bibinfo {author} {\bibfnamefont {F.}~\bibnamefont
  {Drepper}}\ and\ \bibinfo {author} {\bibfnamefont {J.~S.}\ \bibnamefont
  {Briggs}},\ }\bibfield  {title} {\bibinfo {title} {Doubly differential cross
  sections for electron-loss in ion-atom collisions},\ }\href
  {https://doi.org/10.1088/0022-3700/9/12/019} {\bibfield  {journal} {\bibinfo
  {journal} {J. Phys. B}\ }\textbf {\bibinfo {volume} {9}},\ \bibinfo {pages}
  {2063} (\bibinfo {year} {1976})}\BibitemShut {NoStop}%
\bibitem [{\citenamefont {Briggs}\ and\ \citenamefont
  {Drepper}(1978)}]{briggs_asymptotic_1978}%
  \BibitemOpen
  \bibfield  {author} {\bibinfo {author} {\bibfnamefont {J.~S.}\ \bibnamefont
  {Briggs}}\ and\ \bibinfo {author} {\bibfnamefont {F.}~\bibnamefont
  {Drepper}},\ }\bibfield  {title} {\bibinfo {title} {Asymptotic form of the
  cross section for electron loss into the forward direction},\ }\href
  {https://doi.org/10.1088/0022-3700/11/23/013} {\bibfield  {journal} {\bibinfo
   {journal} {J. Phys. B}\ }\textbf {\bibinfo {volume} {11}},\ \bibinfo {pages}
  {4033} (\bibinfo {year} {1978})}\BibitemShut {NoStop}%
\bibitem [{\citenamefont {Briggs}\ and\ \citenamefont
  {Day}(1980)}]{briggs_structure_1980}%
  \BibitemOpen
  \bibfield  {author} {\bibinfo {author} {\bibfnamefont {J.~S.}\ \bibnamefont
  {Briggs}}\ and\ \bibinfo {author} {\bibfnamefont {M.~H.}\ \bibnamefont
  {Day}},\ }\bibfield  {title} {\bibinfo {title} {The structure of the doubly
  differential cross section for electron loss to low-lying continuum states of
  the projectile in fast ion-atom collisions},\ }\href
  {https://doi.org/10.1088/0022-3700/13/24/014} {\bibfield  {journal} {\bibinfo
   {journal} {J. Phys. B}\ }\textbf {\bibinfo {volume} {13}},\ \bibinfo {pages}
  {4797} (\bibinfo {year} {1980})}\BibitemShut {NoStop}%
\bibitem [{\citenamefont {Burgd{\"o}rfer}\ \emph {et~al.}(1983)\citenamefont
  {Burgd{\"o}rfer}, \citenamefont {Breinig}, \citenamefont {Elston},\ and\
  \citenamefont {Sellin}}]{burgdorfer_calculation_1983}%
  \BibitemOpen
  \bibfield  {author} {\bibinfo {author} {\bibfnamefont {J.}~\bibnamefont
  {Burgd{\"o}rfer}}, \bibinfo {author} {\bibfnamefont {M.}~\bibnamefont
  {Breinig}}, \bibinfo {author} {\bibfnamefont {S.~B.}\ \bibnamefont
  {Elston}},\ and\ \bibinfo {author} {\bibfnamefont {I.~A.}\ \bibnamefont
  {Sellin}},\ }\bibfield  {title} {\bibinfo {title} {Calculation of
  electron-loss-to-continuum cusps: {An} algebraic approach},\ }\href
  {https://doi.org/10.1103/PhysRevA.28.3277} {\bibfield  {journal} {\bibinfo
  {journal} {Phys. Rev. A}\ }\textbf {\bibinfo {volume} {28}},\ \bibinfo
  {pages} {3277} (\bibinfo {year} {1983})}\BibitemShut {NoStop}%
\bibitem [{\citenamefont {Elston}\ \emph {et~al.}(1985)\citenamefont {Elston},
  \citenamefont {Berry}, \citenamefont {Burgd{\"o}rfer}, \citenamefont
  {Sellin}, \citenamefont {Breinig}, \citenamefont {DeSerio}, \citenamefont
  {Gonzalez-Lepera}, \citenamefont {Liljeby}, \citenamefont {Groeneveld},
  \citenamefont {Hofmann}, \citenamefont {Koschar},\ and\ \citenamefont
  {Nemirovsky}}]{elston_observation_1985}%
  \BibitemOpen
  \bibfield  {author} {\bibinfo {author} {\bibfnamefont {S.~B.}\ \bibnamefont
  {Elston}}, \bibinfo {author} {\bibfnamefont {S.~D.}\ \bibnamefont {Berry}},
  \bibinfo {author} {\bibfnamefont {J.}~\bibnamefont {Burgd{\"o}rfer}},
  \bibinfo {author} {\bibfnamefont {I.~A.}\ \bibnamefont {Sellin}}, \bibinfo
  {author} {\bibfnamefont {M.}~\bibnamefont {Breinig}}, \bibinfo {author}
  {\bibfnamefont {R.}~\bibnamefont {DeSerio}}, \bibinfo {author} {\bibfnamefont
  {C.~E.}\ \bibnamefont {Gonzalez-Lepera}}, \bibinfo {author} {\bibfnamefont
  {L.}~\bibnamefont {Liljeby}}, \bibinfo {author} {\bibfnamefont {K.~O.}\
  \bibnamefont {Groeneveld}}, \bibinfo {author} {\bibfnamefont
  {D.}~\bibnamefont {Hofmann}}, \emph {et~al.},\ }\bibfield  {title} {\bibinfo
  {title} {Observation of {Quadrupole} and {Hexadecapole} {Moments} of the
  {Electronic} {Charge} {Cloud} {Produced} in {Electron}-{Loss} {Collisions}},\
  }\href {https://doi.org/10.1103/PhysRevLett.55.2281} {\bibfield  {journal}
  {\bibinfo  {journal} {Phys. Rev. Lett.}\ }\textbf {\bibinfo {volume} {55}},\
  \bibinfo {pages} {2281} (\bibinfo {year} {1985})}\BibitemShut {NoStop}%
\bibitem [{\citenamefont {Oswald}\ \emph {et~al.}(1989)\citenamefont {Oswald},
  \citenamefont {Schramm},\ and\ \citenamefont
  {Betz}}]{oswald_higher-order_1989}%
  \BibitemOpen
  \bibfield  {author} {\bibinfo {author} {\bibfnamefont {W.}~\bibnamefont
  {Oswald}}, \bibinfo {author} {\bibfnamefont {R.}~\bibnamefont {Schramm}},\
  and\ \bibinfo {author} {\bibfnamefont {H.-D.}\ \bibnamefont {Betz}},\
  }\bibfield  {title} {\bibinfo {title} {Higher-order effects of electron
  capture and loss to the continuum in ion-atom collisions},\ }\href
  {https://doi.org/10.1103/PhysRevLett.62.1114} {\bibfield  {journal} {\bibinfo
   {journal} {Phys. Rev. Lett.}\ }\textbf {\bibinfo {volume} {62}},\ \bibinfo
  {pages} {1114} (\bibinfo {year} {1989})}\BibitemShut {NoStop}%
\bibitem [{\citenamefont {Atan}\ \emph {et~al.}(1990)\citenamefont {Atan},
  \citenamefont {Steckelmacher},\ and\ \citenamefont
  {Lucas}}]{atan_multipole_1990}%
  \BibitemOpen
  \bibfield  {author} {\bibinfo {author} {\bibfnamefont {H.}~\bibnamefont
  {Atan}}, \bibinfo {author} {\bibfnamefont {W.}~\bibnamefont
  {Steckelmacher}},\ and\ \bibinfo {author} {\bibfnamefont {M.~W.}\
  \bibnamefont {Lucas}},\ }\bibfield  {title} {\bibinfo {title} {Multipole
  expansion analysis of electron loss to the continuum for {He$^+$} ions colliding
  with rare gases},\ }\href {https://doi.org/10.1088/0953-4075/23/15/024}
  {\bibfield  {journal} {\bibinfo  {journal} {J. Phys. B: At. Mol. Opt. Phys.}\
  }\textbf {\bibinfo {volume} {23}},\ \bibinfo {pages} {2579} (\bibinfo {year}
  {1990})}\BibitemShut {NoStop}%
\bibitem [{\citenamefont
  {Jakubassa-Amundsen}(1990)}]{jakubassa-amundsen_second_1990}%
  \BibitemOpen
  \bibfield  {author} {\bibinfo {author} {\bibfnamefont {D.~H.}\ \bibnamefont
  {Jakubassa-Amundsen}},\ }\bibfield  {title} {\bibinfo {title} {The second
  {Born} approximation for electron loss to the continuum},\ }\href
  {https://doi.org/10.1088/0953-4075/23/19/017} {\bibfield  {journal} {\bibinfo
   {journal} {J. Phys. B}\ }\textbf {\bibinfo {volume} {23}},\ \bibinfo {pages}
  {3335} (\bibinfo {year} {1990})}\BibitemShut {NoStop}%
\bibitem [{\citenamefont {Guly{\'a}s}\ \emph {et~al.}(1992)\citenamefont
  {Guly{\'a}s}, \citenamefont {Sarkadi}, \citenamefont {P{\'a}link{\'a}s},
  \citenamefont {K{\"o}v{\'e}r}, \citenamefont {Vajnai}, \citenamefont
  {Szab{\'o}}, \citenamefont {V{\'e}gh}, \citenamefont {Ber{\'e}nyi},\ and\
  \citenamefont {B.~Elston}}]{gulyas_cusp-shape_1992}%
  \BibitemOpen
  \bibfield  {author} {\bibinfo {author} {\bibfnamefont {L.}~\bibnamefont
  {Guly{\'a}s}}, \bibinfo {author} {\bibfnamefont {L.}~\bibnamefont {Sarkadi}},
  \bibinfo {author} {\bibfnamefont {J.}~\bibnamefont {P{\'a}link{\'a}s}},
  \bibinfo {author} {\bibfnamefont {{\'A}.}~\bibnamefont {K{\"o}v{\'e}r}},
  \bibinfo {author} {\bibfnamefont {T.}~\bibnamefont {Vajnai}}, \bibinfo
  {author} {\bibfnamefont {G.}~\bibnamefont {Szab{\'o}}}, \bibinfo {author}
  {\bibfnamefont {J.}~\bibnamefont {V{\'e}gh}}, \bibinfo {author}
  {\bibfnamefont {D.}~\bibnamefont {Ber{\'e}nyi}},\ and\ \bibinfo {author}
  {\bibfnamefont {S.}~\bibnamefont {B.~Elston}},\ }\bibfield  {title} {\bibinfo
  {title} {Cusp-shape studies with {He$^+$} ions at 1.41- and 2.41-a.u. impact
  velocities},\ }\href {https://doi.org/10.1103/PhysRevA.45.4535} {\bibfield
  {journal} {\bibinfo  {journal} {Phys. Rev. A}\ }\textbf {\bibinfo {volume}
  {45}},\ \bibinfo {pages} {4535} (\bibinfo {year} {1992})}\BibitemShut
  {NoStop}%
\bibitem [{\citenamefont
  {Jakubassa-Amundsen}(1993)}]{jakubassa-amundsen_strong_1993}%
  \BibitemOpen
  \bibfield  {author} {\bibinfo {author} {\bibfnamefont {D.~H.}\ \bibnamefont
  {Jakubassa-Amundsen}},\ }\bibfield  {title} {\bibinfo {title} {Strong
  potential second {Born} theory for electron loss to the continuum in
  collision with heavy targets},\ }\href
  {https://doi.org/10.1088/0953-4075/26/17/018} {\bibfield  {journal} {\bibinfo
   {journal} {J. Phys. B}\ }\textbf {\bibinfo {volume} {26}},\ \bibinfo {pages}
  {2853} (\bibinfo {year} {1993})}\BibitemShut {NoStop}%
\bibitem [{\citenamefont {Hillenbrand}\ \emph {et~al.}(2016)\citenamefont
  {Hillenbrand}, \citenamefont {Hagmann}, \citenamefont {Monti}, \citenamefont
  {Rivarola}, \citenamefont {Blumenhagen}, \citenamefont {Brandau},
  \citenamefont {Chen}, \citenamefont {DuBois}, \citenamefont {Gumberidze},
  \citenamefont {Guo}, \citenamefont {Lestinsky}, \citenamefont {Litvinov},
  \citenamefont {M{\"u}ller}, \citenamefont {Schippers}, \citenamefont
  {Spillmann}, \citenamefont {Trotsenko}, \citenamefont {Weber},\ and\
  \citenamefont {St{\"o}hlker}}]{hillenbrand_strong_2016}%
  \BibitemOpen
  \bibfield  {author} {\bibinfo {author} {\bibfnamefont {P.-M.}\ \bibnamefont
  {Hillenbrand}}, \bibinfo {author} {\bibfnamefont {S.}~\bibnamefont
  {Hagmann}}, \bibinfo {author} {\bibfnamefont {J.~M.}\ \bibnamefont {Monti}},
  \bibinfo {author} {\bibfnamefont {R.~D.}\ \bibnamefont {Rivarola}}, \bibinfo
  {author} {\bibfnamefont {K.-H.}\ \bibnamefont {Blumenhagen}}, \bibinfo
  {author} {\bibfnamefont {C.}~\bibnamefont {Brandau}}, \bibinfo {author}
  {\bibfnamefont {W.}~\bibnamefont {Chen}}, \bibinfo {author} {\bibfnamefont
  {R.~D.}\ \bibnamefont {DuBois}}, \bibinfo {author} {\bibfnamefont
  {A.}~\bibnamefont {Gumberidze}}, \bibinfo {author} {\bibfnamefont {D.~L.}\
  \bibnamefont {Guo}}, \emph {et~al.},\  }\bibfield  {title} {\bibinfo {title} {Strong asymmetry of the
  electron-loss-to-continuum cusp of multielectron {U$^{28+}$} projectiles in
  near-relativistic collisions with gaseous targets},\ }\href
  {https://doi.org/10.1103/PhysRevA.93.042709} {\bibfield  {journal} {\bibinfo
  {journal} {Phys. Rev. A}\ }\textbf {\bibinfo {volume} {93}},\ \bibinfo
  {pages} {042709} (\bibinfo {year} {2016})}\BibitemShut {NoStop}%
\bibitem [{\citenamefont {Hillenbrand}\ \emph
  {et~al.}(2014{\natexlab{a}})\citenamefont {Hillenbrand}, \citenamefont
  {Hagmann}, \citenamefont {Voitkiv}, \citenamefont {Najjari}, \citenamefont
  {Bana{\'s}}, \citenamefont {Blumenhagen}, \citenamefont {Brandau}, \citenamefont
  {Chen}, \citenamefont {De~Filippo}, \citenamefont {Gumberidze}, \citenamefont
  {Guo}, \citenamefont {Kozhuharov}, \citenamefont {Lestinsky}, \citenamefont
  {Litvinov}, \citenamefont {M{\"u}ller}, \citenamefont {Rothard},
  \citenamefont {Schippers}, \citenamefont {Sch{\"o}ffler}, \citenamefont
  {Spillmann}, \citenamefont {Trotsenko}, \citenamefont {Zhu},\ and\
  \citenamefont {St{\"o}hlker}}]{hillenbrand_electron-loss--continuum_2014}%
  \BibitemOpen
  \bibfield  {author} {\bibinfo {author} {\bibfnamefont {P.-M.}\ \bibnamefont
  {Hillenbrand}}, \bibinfo {author} {\bibfnamefont {S.}~\bibnamefont
  {Hagmann}}, \bibinfo {author} {\bibfnamefont {A.~B.}\ \bibnamefont
  {Voitkiv}}, \bibinfo {author} {\bibfnamefont {B.}~\bibnamefont {Najjari}},
  \bibinfo {author} {\bibfnamefont {D.}~\bibnamefont {Bana{\'s}}}, \bibinfo
  {author} {\bibfnamefont {K.-H.}\ \bibnamefont {Blumenhagen}}, \bibinfo
  {author} {\bibfnamefont {C.}~\bibnamefont {Brandau}}, \bibinfo {author}
  {\bibfnamefont {W.}~\bibnamefont {Chen}}, \bibinfo {author} {\bibfnamefont
  {E.}~\bibnamefont {De~Filippo}}, \bibinfo {author} {\bibfnamefont
  {A.}~\bibnamefont {Gumberidze}}, \emph {et~al.},\ }\bibfield  {title}
  {\bibinfo {title} {Electron-loss-to-continuum cusp in {U$^{88+}$} + {N$_2$}
  collisions},\ }\href {https://doi.org/10.1103/PhysRevA.90.042713} {\bibfield
  {journal} {\bibinfo  {journal} {Phys. Rev. A}\ }\textbf {\bibinfo {volume}
  {90}},\ \bibinfo {pages} {042713} (\bibinfo {year}
  {2014}{\natexlab{a}})}\BibitemShut {NoStop}%
\bibitem [{\citenamefont {Hillenbrand}\ \emph {et~al.}(2020)\citenamefont
  {Hillenbrand}, \citenamefont {Hagmann}, \citenamefont {Groshev},
  \citenamefont {Bana{\'s}}, \citenamefont {Benis}, \citenamefont {Brandau},
  \citenamefont {De~Filippo}, \citenamefont {Forstner}, \citenamefont
  {Glorius}, \citenamefont {Grisenti}, \citenamefont {Gumberidze},
  \citenamefont {Guo}, \citenamefont {Hai}, \citenamefont {Herdrich},
  \citenamefont {Lestinsky}, \citenamefont {Litvinov}, \citenamefont {Pagano},
  \citenamefont {Petridis}, \citenamefont {Sanjari}, \citenamefont {Schury},
  \citenamefont {Spillmann}, \citenamefont {Trotsenko}, \citenamefont
  {Vockert}, \citenamefont {Weber}, \citenamefont {Yerokhin},\ and\
  \citenamefont {St{\"o}hlker}}]{hillenbrand_radiative_2020}%
  \BibitemOpen
  \bibfield  {author} {\bibinfo {author} {\bibfnamefont {P.-M.}\ \bibnamefont
  {Hillenbrand}}, \bibinfo {author} {\bibfnamefont {S.}~\bibnamefont
  {Hagmann}}, \bibinfo {author} {\bibfnamefont {M.~E.}\ \bibnamefont
  {Groshev}}, \bibinfo {author} {\bibfnamefont {D.}~\bibnamefont {Bana{\'s}}},
  \bibinfo {author} {\bibfnamefont {E.~P.}\ \bibnamefont {Benis}}, \bibinfo
  {author} {\bibfnamefont {C.}~\bibnamefont {Brandau}}, \bibinfo {author}
  {\bibfnamefont {E.}~\bibnamefont {De~Filippo}}, \bibinfo {author}
  {\bibfnamefont {O.}~\bibnamefont {Forstner}}, \bibinfo {author}
  {\bibfnamefont {J.}~\bibnamefont {Glorius}}, \bibinfo {author} {\bibfnamefont
  {R.~E.}\ \bibnamefont {Grisenti}}, \emph {et~al.},\ }\bibfield  {title} {\bibinfo {title}
  {Radiative electron capture to the continuum in {U$^{89+}$} + {N$_2$} collisions:
  {Experiment} and theory},\ }\href
  {https://doi.org/10.1103/PhysRevA.101.022708} {\bibfield  {journal} {\bibinfo
   {journal} {Phys. Rev. A}\ }\textbf {\bibinfo {volume} {101}},\ \bibinfo
  {pages} {022708} (\bibinfo {year} {2020})}\BibitemShut {NoStop}%
\bibitem [{\citenamefont {Voitkiv}(2007)}]{voitkiv_relativistic_2007}%
  \BibitemOpen
  \bibfield  {author} {\bibinfo {author} {\bibfnamefont {A.~B.}\ \bibnamefont
  {Voitkiv}},\ }\bibfield  {title} {\bibinfo {title} {On the relativistic and
  nonrelativistic electron descriptions in high-energy atomic collisions},\
  }\href {https://doi.org/10.1088/0953-4075/40/14/008} {\bibfield  {journal}
  {\bibinfo  {journal} {J. Phys. B}\ }\textbf {\bibinfo {volume} {40}},\
  \bibinfo {pages} {2885} (\bibinfo {year} {2007})}\BibitemShut {NoStop}%
\bibitem [{\citenamefont {Voitkiv}\ and\ \citenamefont
  {Ullrich}(2008)}]{voitkiv_relativistic_2008}%
  \BibitemOpen
  \bibfield  {author} {\bibinfo {author} {\bibfnamefont {A.}~\bibnamefont
  {Voitkiv}}\ and\ \bibinfo {author} {\bibfnamefont {J.}~\bibnamefont
  {Ullrich}},\ }\href
  {http://link.springer.com/book/10.1007%2F978-3-540-78421-0} {\emph {\bibinfo
  {title} {Relativistic {Collisions} of {Structured} {Atomic} {Particles}}}},\
  \bibinfo {series} {Springer {Series} on {Atomic}, {Optical}, and {Plasma}
  {Physics}}\ No.~\bibinfo {number} {49}\ (\bibinfo  {publisher} {Springer},\
  \bibinfo {year} {2008})\BibitemShut {NoStop}%
\bibitem [{\citenamefont {Momberger}\ \emph {et~al.}(1989)\citenamefont
  {Momberger}, \citenamefont {Gr{\"u}n}, \citenamefont {Scheid},\ and\
  \citenamefont {Becker}}]{momberger_angular_1989}%
  \BibitemOpen
  \bibfield  {author} {\bibinfo {author} {\bibfnamefont {K.}~\bibnamefont
  {Momberger}}, \bibinfo {author} {\bibfnamefont {N.}~\bibnamefont {Gr{\"u}n}},
  \bibinfo {author} {\bibfnamefont {W.}~\bibnamefont {Scheid}},\ and\ \bibinfo
  {author} {\bibfnamefont {U.}~\bibnamefont {Becker}},\ }\bibfield  {title}
  {\bibinfo {title} {Angular distribution of electrons emitted from $1s_{1/2}$ and
  $2s_{1/2}$ states in relativistic heavy-ion collisions},\ }\href
  {https://doi.org/10.1088/0953-4075/22/20/018} {\bibfield  {journal} {\bibinfo
   {journal} {J. Phys. B}\ }\textbf {\bibinfo {volume} {22}},\ \bibinfo {pages}
  {3269} (\bibinfo {year} {1989})}\BibitemShut {NoStop}%
\bibitem [{\citenamefont {Voitkiv}\ \emph {et~al.}(2000)\citenamefont
  {Voitkiv}, \citenamefont {Gr{\"u}n},\ and\ \citenamefont
  {Scheid}}]{voitkiv_plane-wave_2000}%
  \BibitemOpen
  \bibfield  {author} {\bibinfo {author} {\bibfnamefont {A.~B.}\ \bibnamefont
  {Voitkiv}}, \bibinfo {author} {\bibfnamefont {N.}~\bibnamefont {Gr{\"u}n}},\
  and\ \bibinfo {author} {\bibfnamefont {W.}~\bibnamefont {Scheid}},\
  }\bibfield  {title} {\bibinfo {title} {Plane-wave {Born} treatment of
  projectile-electron excitation and loss in relativistic collisions with
  atomic targets},\ }\href {https://doi.org/10.1103/PhysRevA.61.052704}
  {\bibfield  {journal} {\bibinfo  {journal} {Phys. Rev. A}\ }\textbf {\bibinfo
  {volume} {61}},\ \bibinfo {pages} {052704} (\bibinfo {year}
  {2000})}\BibitemShut {NoStop}%
\bibitem [{\citenamefont {Surzhykov}\ and\ \citenamefont
  {Fritzsche}(2005)}]{surzhykov_electron_2005}%
  \BibitemOpen
  \bibfield  {author} {\bibinfo {author} {\bibfnamefont {A.}~\bibnamefont
  {Surzhykov}}\ and\ \bibinfo {author} {\bibfnamefont {S.}~\bibnamefont
  {Fritzsche}},\ }\bibfield  {title} {\bibinfo {title} {Electron angular and
  energy distributions following the ionization of highly charged projectile
  ions},\ }\href {https://doi.org/10.1088/0953-4075/38/15/011} {\bibfield
  {journal} {\bibinfo  {journal} {J. Phys. B}\ }\textbf {\bibinfo {volume}
  {38}},\ \bibinfo {pages} {2711} (\bibinfo {year} {2005})}\BibitemShut
  {NoStop}%
\bibitem [{\citenamefont {Lyashchenko}\ \emph {et~al.}(2018)\citenamefont
  {Lyashchenko}, \citenamefont {Andreev},\ and\ \citenamefont
  {Voitkiv}}]{lyashchenko_electron_2018}%
  \BibitemOpen
  \bibfield  {author} {\bibinfo {author} {\bibfnamefont {K.~N.}\ \bibnamefont
  {Lyashchenko}}, \bibinfo {author} {\bibfnamefont {O.~Y.}\ \bibnamefont
  {Andreev}},\ and\ \bibinfo {author} {\bibfnamefont {A.~B.}\ \bibnamefont
  {Voitkiv}},\ }\bibfield  {title} {\bibinfo {title} {Electron loss from
  hydrogen-like highly charged ions in collisions with electrons, protons and
  light atoms},\ }\href {https://doi.org/10.1088/1361-6455/aaaa11} {\bibfield
  {journal} {\bibinfo  {journal} {J. Phys. B}\ }\textbf {\bibinfo {volume}
  {51}},\ \bibinfo {pages} {055204} (\bibinfo {year} {2018})}\BibitemShut
  {NoStop}%
\bibitem [{\citenamefont {Bondarev}\ \emph {et~al.}(2020)\citenamefont
  {Bondarev}, \citenamefont {Kozhedub}, \citenamefont {Tupitsyn}, \citenamefont
  {Shabaev},\ and\ \citenamefont {Plunien}}]{bondarev_calculations_2020}%
  \BibitemOpen
  \bibfield  {author} {\bibinfo {author} {\bibfnamefont {A.~I.}\ \bibnamefont
  {Bondarev}}, \bibinfo {author} {\bibfnamefont {Y.~S.}\ \bibnamefont
  {Kozhedub}}, \bibinfo {author} {\bibfnamefont {I.~I.}\ \bibnamefont
  {Tupitsyn}}, \bibinfo {author} {\bibfnamefont {V.~M.}\ \bibnamefont
  {Shabaev}},\ and\ \bibinfo {author} {\bibfnamefont {G.}~\bibnamefont
  {Plunien}},\ }\bibfield  {title} {\bibinfo {title} {Calculations of
  {Electron} {Loss} to {Continuum} in {Collisions} of {Li}- and {Be}-{Like}
  {Uranium} {Ions} with {Nitrogen} {Targets}},\ }\href
  {https://doi.org/10.3390/atoms8040089} {\bibfield  {journal} {\bibinfo
  {journal} {Atoms}\ }\textbf {\bibinfo {volume} {8}},\ \bibinfo {pages} {89}
  (\bibinfo {year} {2020})}\BibitemShut {NoStop}%
\bibitem [{\citenamefont {Belkic}(1978)}]{belkic_quantum_1978}%
  \BibitemOpen
  \bibfield  {author} {\bibinfo {author} {\bibfnamefont {D.}~\bibnamefont
  {Belki{\'c}}},\ }\bibfield  {title} {\bibinfo {title} {A quantum theory of
  ionisation in fast collisions between ions and atomic systems},\ }\href
  {https://doi.org/10.1088/0022-3700/11/20/015} {\bibfield  {journal} {\bibinfo
   {journal} {J. Phys. B}\ }\textbf {\bibinfo {volume} {11}},\ \bibinfo {pages}
  {3529} (\bibinfo {year} {1978})}\BibitemShut {NoStop}%
\bibitem [{\citenamefont {Belkic}(1979)}]{belkic_electron_1979}%
  \BibitemOpen
  \bibfield  {author} {\bibinfo {author} {\bibfnamefont {D.}~\bibnamefont
  {Belki{\'c}}},\ }\bibfield  {title} {\bibinfo {title} {Electron capture in
  high-energy ion-atom collisions},\ }\href
  {https://doi.org/10.1016/0370-1573(79)90035-8} {\bibfield  {journal}
  {\bibinfo  {journal} {Phys. Rep.}\ }\textbf {\bibinfo {volume} {56}},\
  \bibinfo {pages} {279} (\bibinfo {year} {1979})}\BibitemShut {NoStop}%
\bibitem [{\citenamefont {Crothers}\ and\ \citenamefont
  {McCann}(1983)}]{crothers_ionisation_1983}%
  \BibitemOpen
  \bibfield  {author} {\bibinfo {author} {\bibfnamefont {D.~S.~F.}\
  \bibnamefont {Crothers}}\ and\ \bibinfo {author} {\bibfnamefont {J.~F.}\
  \bibnamefont {McCann}},\ }\bibfield  {title} {\bibinfo {title} {Ionisation of
  atoms by ion impact},\ }\href {https://doi.org/10.1088/0022-3700/16/17/015}
  {\bibfield  {journal} {\bibinfo  {journal} {J. Phys. B}\ }\textbf {\bibinfo
  {volume} {16}},\ \bibinfo {pages} {3229} (\bibinfo {year}
  {1983})}\BibitemShut {NoStop}%
\bibitem [{\citenamefont {Maidagan}\ and\ \citenamefont
  {Rivarola}(1984)}]{maidagan_symmetric_1984}%
  \BibitemOpen
  \bibfield  {author} {\bibinfo {author} {\bibfnamefont {J.~M.}\ \bibnamefont
  {Maidagan}}\ and\ \bibinfo {author} {\bibfnamefont {R.~D.}\ \bibnamefont
  {Rivarola}},\ }\bibfield  {title} {\bibinfo {title} {A symmetric eikonal-type
  approximation for electron capture in ion-atom collisions},\ }\href
  {https://doi.org/10.1088/0022-3700/17/12/016} {\bibfield  {journal} {\bibinfo
   {journal} {J. Phys. B}\ }\textbf {\bibinfo {volume} {17}},\ \bibinfo {pages}
  {2477} (\bibinfo {year} {1984})}\BibitemShut {NoStop}%
\bibitem [{\citenamefont {Deco}\ \emph {et~al.}(1986)\citenamefont {Deco},
  \citenamefont {Fainstein},\ and\ \citenamefont
  {Rivarola}}]{deco_symmetric_1986}%
  \BibitemOpen
  \bibfield  {author} {\bibinfo {author} {\bibfnamefont {G.~R.}\ \bibnamefont
  {Deco}}, \bibinfo {author} {\bibfnamefont {P.~D.}\ \bibnamefont
  {Fainstein}},\ and\ \bibinfo {author} {\bibfnamefont {R.~D.}\ \bibnamefont
  {Rivarola}},\ }\bibfield  {title} {\bibinfo {title} {Symmetric eikonal
  approximation for electron excitation in ion-atom collisions},\ }\href
  {https://doi.org/10.1088/0022-3700/19/2/010} {\bibfield  {journal} {\bibinfo
  {journal} {J. Phys. B}\ }\textbf {\bibinfo {volume} {19}},\ \bibinfo {pages}
  {213} (\bibinfo {year} {1986})}\BibitemShut {NoStop}%
\bibitem [{\citenamefont {Fainstein}\ and\ \citenamefont
  {Rivarola}(1987)}]{fainstein_symmetric_1987}%
  \BibitemOpen
  \bibfield  {author} {\bibinfo {author} {\bibfnamefont {P.~D.}\ \bibnamefont
  {Fainstein}}\ and\ \bibinfo {author} {\bibfnamefont {R.~D.}\ \bibnamefont
  {Rivarola}},\ }\bibfield  {title} {\bibinfo {title} {Symmetric eikonal model
  for ionisation in ion-atom collisions},\ }\href
  {https://doi.org/10.1088/0022-3700/20/6/015} {\bibfield  {journal} {\bibinfo
  {journal} {J. Phys. B}\ }\textbf {\bibinfo {volume} {20}},\ \bibinfo {pages}
  {1285} (\bibinfo {year} {1987})}\BibitemShut {NoStop}%
\bibitem [{\citenamefont {Martinez}\ and\ \citenamefont
  {Rivarola}(1990)}]{martinez_second-order_1990}%
  \BibitemOpen
  \bibfield  {author} {\bibinfo {author} {\bibfnamefont {A.~E.}\ \bibnamefont
  {Martinez}}\ and\ \bibinfo {author} {\bibfnamefont {R.~D.}\ \bibnamefont
  {Rivarola}},\ }\bibfield  {title} {\bibinfo {title} {Second-order
  distorted-wave approximations for charge exchange},\ }\href
  {https://doi.org/10.1088/0953-4075/23/22/017} {\bibfield  {journal} {\bibinfo
   {journal} {J. Phys. B}\ }\textbf {\bibinfo {volume} {23}},\ \bibinfo {pages}
  {4165} (\bibinfo {year} {1990})}\BibitemShut {NoStop}%
\bibitem [{\citenamefont {Fainstein}\ \emph {et~al.}(1991)\citenamefont
  {Fainstein}, \citenamefont {Ponce},\ and\ \citenamefont
  {Rivarola}}]{fainstein_two-centre_1991}%
  \BibitemOpen
  \bibfield  {author} {\bibinfo {author} {\bibfnamefont {P.~D.}\ \bibnamefont
  {Fainstein}}, \bibinfo {author} {\bibfnamefont {V.~H.}\ \bibnamefont
  {Ponce}},\ and\ \bibinfo {author} {\bibfnamefont {R.~D.}\ \bibnamefont
  {Rivarola}},\ }\bibfield  {title} {\bibinfo {title} {Two-centre effects in
  ionization by ion impact},\ }\href
  {https://doi.org/10.1088/0953-4075/24/14/005} {\bibfield  {journal} {\bibinfo
   {journal} {J. Phys. B}\ }\textbf {\bibinfo {volume} {24}},\ \bibinfo {pages}
  {3091} (\bibinfo {year} {1991})}\BibitemShut {NoStop}%
\bibitem [{\citenamefont {Dewangan}\ and\ \citenamefont
  {Eichler}(1994)}]{dewangan_charge_1994}%
  \BibitemOpen
  \bibfield  {author} {\bibinfo {author} {\bibfnamefont {D.}~\bibnamefont
  {Dewangan}}\ and\ \bibinfo {author} {\bibfnamefont {J.}~\bibnamefont
  {Eichler}},\ }\bibfield  {title} {\bibinfo {title} {Charge exchange in
  energetic ion-atom collisions},\ }\href
  {https://doi.org/10.1016/0370-1573(94)90012-4} {\bibfield  {journal}
  {\bibinfo  {journal} {Phys. Rep.}\ }\textbf {\bibinfo {volume} {247}},\
  \bibinfo {pages} {59} (\bibinfo {year} {1994})}\BibitemShut {NoStop}%
\bibitem [{\citenamefont {Toshima}\ and\ \citenamefont
  {Eichler}(1990)}]{toshima_distorted-wave_1990}%
  \BibitemOpen
  \bibfield  {author} {\bibinfo {author} {\bibfnamefont {N.}~\bibnamefont
  {Toshima}}\ and\ \bibinfo {author} {\bibfnamefont {J.}~\bibnamefont
  {Eichler}},\ }\bibfield  {title} {\bibinfo {title} {Distorted-wave
  approximations for relativistic atomic collisions},\ }\href
  {https://doi.org/10.1103/PhysRevA.41.5221} {\bibfield  {journal} {\bibinfo
  {journal} {Phy. Rev. A}\ }\textbf {\bibinfo {volume} {41}},\ \bibinfo {pages}
  {5221} (\bibinfo {year} {1990})}\BibitemShut {NoStop}%
\bibitem [{\citenamefont {Crothers}(2000)}]{crothers_relativistic_2000}%
  \BibitemOpen
  \bibfield  {author} {\bibinfo {author} {\bibfnamefont {D.~S.~F.}\
  \bibnamefont {Crothers}},\ }\href {https://doi.org/10.1007/978-1-4615-4275-9}
  {\emph {\bibinfo {title} {Relativistic {Heavy}-{Particle} {Collision}
  {Theory}}}}\ (\bibinfo  {publisher} {Springer US},\ \bibinfo {address}
  {Boston, MA},\ \bibinfo {year} {2000})\BibitemShut {NoStop}%
\bibitem [{\citenamefont {Voitkiv}\ and\ \citenamefont
  {Najjari}(2007)}]{voitkiv_three-body_2007}%
  \BibitemOpen
  \bibfield  {author} {\bibinfo {author} {\bibfnamefont {A.~B.}\ \bibnamefont
  {Voitkiv}}\ and\ \bibinfo {author} {\bibfnamefont {B.}~\bibnamefont
  {Najjari}},\ }\bibfield  {title} {\bibinfo {title} {Three-body models for the
  electron loss from the $K$-shell of highly charged ions in relativistic
  collisions with atoms},\ }\href {https://doi.org/10.1088/0953-4075/40/16/010}
  {\bibfield  {journal} {\bibinfo  {journal} {J. Phys. B}\ }\textbf {\bibinfo
  {volume} {40}},\ \bibinfo {pages} {3295} (\bibinfo {year}
  {2007})}\BibitemShut {NoStop}%
\bibitem [{\citenamefont {Voitkiv}\ \emph {et~al.}(2007)\citenamefont
  {Voitkiv}, \citenamefont {Najjari},\ and\ \citenamefont
  {Ullrich}}]{voitkiv_electron_2007}%
  \BibitemOpen
  \bibfield  {author} {\bibinfo {author} {\bibfnamefont {A.~B.}\ \bibnamefont
  {Voitkiv}}, \bibinfo {author} {\bibfnamefont {B.}~\bibnamefont {Najjari}},\
  and\ \bibinfo {author} {\bibfnamefont {J.}~\bibnamefont {Ullrich}},\
  }\bibfield  {title} {\bibinfo {title} {Electron loss from hydrogenlike,
  heliumlike, and lithiumlike uranium ions in collisions with atoms at low
  relativistic impact energies},\ }\href
  {https://doi.org/10.1103/PhysRevA.76.022709} {\bibfield  {journal} {\bibinfo
  {journal} {Phys. Rev. A}\ }\textbf {\bibinfo {volume} {76}},\ \bibinfo
  {pages} {022709} (\bibinfo {year} {2007})}\BibitemShut {NoStop}%
\bibitem [{\citenamefont {Hillenbrand}\ \emph
  {et~al.}(2014{\natexlab{b}})\citenamefont {Hillenbrand}, \citenamefont
  {Hagmann}, \citenamefont {Atanasov}, \citenamefont {Bana{\'s}}, \citenamefont
  {Blumenhagen}, \citenamefont {Brandau}, \citenamefont {Chen}, \citenamefont
  {De~Filippo}, \citenamefont {Gumberidze}, \citenamefont {Guo}, \citenamefont
  {Jakubassa-Amundsen}, \citenamefont {Kovtun}, \citenamefont {Kozhuharov},
  \citenamefont {Lestinsky}, \citenamefont {Litvinov}, \citenamefont
  {M{\"u}ller}, \citenamefont {M{\"u}ller}, \citenamefont {Rothard},
  \citenamefont {Schippers}, \citenamefont {Sch{\"o}ffler}, \citenamefont
  {Spillmann}, \citenamefont {Surzhykov}, \citenamefont {Trotsenko},
  \citenamefont {Winckler}, \citenamefont {Yan}, \citenamefont {Yerokhin},
  \citenamefont {Zhu},\ and\ \citenamefont
  {St{\"o}hlker}}]{hillenbrand_radiative-electron-capture--continuum_2014}%
  \BibitemOpen
  \bibfield  {author} {\bibinfo {author} {\bibfnamefont {P.-M.}\ \bibnamefont
  {Hillenbrand}}, \bibinfo {author} {\bibfnamefont {S.}~\bibnamefont
  {Hagmann}}, \bibinfo {author} {\bibfnamefont {D.}~\bibnamefont {Atanasov}},
  \bibinfo {author} {\bibfnamefont {D.}~\bibnamefont {Bana{\'s}}}, \bibinfo
  {author} {\bibfnamefont {K.-H.}\ \bibnamefont {Blumenhagen}}, \bibinfo
  {author} {\bibfnamefont {C.}~\bibnamefont {Brandau}}, \bibinfo {author}
  {\bibfnamefont {W.}~\bibnamefont {Chen}}, \bibinfo {author} {\bibfnamefont
  {E.}~\bibnamefont {De~Filippo}}, \bibinfo {author} {\bibfnamefont
  {A.}~\bibnamefont {Gumberidze}}, \bibinfo {author} {\bibfnamefont {D.~L.}\
  \bibnamefont {Guo}}, \emph {et~al.},\
  }\bibfield  {title} {\bibinfo {title}
  {Radiative-electron-capture-to-continuum cusp in {U$^{88+}$} +
  {N$_2$} collisions and the high-energy endpoint of electron-nucleus
  bremsstrahlung},\ }\href {https://doi.org/10.1103/PhysRevA.90.022707}
  {\bibfield  {journal} {\bibinfo  {journal} {Phys. Rev. A}\ }\textbf {\bibinfo
  {volume} {90}},\ \bibinfo {pages} {022707} (\bibinfo {year}
  {2014}{\natexlab{b}})}\BibitemShut {NoStop}%
\bibitem [{\citenamefont {Moli\'ere}(1947)}]{moliere1947Naturforsch133}%
  \BibitemOpen
  \bibfield  {author} {\bibinfo {author} {\bibfnamefont {G.}~\bibnamefont
  {Moli\'ere}},\ } \bibfield  {title} 
  {\bibinfo {title} {Theorie  der  Streuung  schneller  geladener  Teilchen I},\ }  
  \href@noop {} {\bibfield  {journal} {\bibinfo  {journal} {Z.
  Naturforsch.}\ }\textbf {\bibinfo {volume} {2A}},\ \bibinfo {pages} {133}
  (\bibinfo {year} {1947})}\BibitemShut {NoStop}%
\bibitem [{\citenamefont {Salvat}\ \emph {et~al.}(1987)\citenamefont {Salvat},
  \citenamefont {Martinez}, \citenamefont {Mayol},\ and\ \citenamefont
  {Parellada}}]{salvat1987pra467}%
  \BibitemOpen
  \bibfield  {author} {\bibinfo {author} {\bibfnamefont {F.}~\bibnamefont
  {Salvat}}, \bibinfo {author} {\bibfnamefont {J.~D.}\ \bibnamefont {Martinez}},
  \bibinfo {author} {\bibfnamefont {R.}~\bibnamefont {Mayol}}, \ and\ \bibinfo
  {author} {\bibfnamefont {J.}~\bibnamefont {Parellada}},\ }\bibfield  {title} 
  {\bibinfo {title} {Analytical Dirac-Hartree-Fock-Slater screening function for atoms (Z=1–92)},\ }
  \href{https://doi.org/10.1103/PhysRevA.36.467}
  {\bibfield  {journal} {\bibinfo  {journal} {Phys. Rev. A}\ }\textbf {\bibinfo
  {volume} {36}},\ \bibinfo {pages} {467} (\bibinfo {year} {1987})}\BibitemShut
  {NoStop}%
 \bibitem [{\citenamefont {Deco}\ and\ \citenamefont
  {Gr{\"u}n}(1989)}]{deco_asymptotic_1989}%
  \BibitemOpen
  \bibfield  {author} {\bibinfo {author} {\bibfnamefont {G.}~\bibnamefont
  {Deco}}\ and\ \bibinfo {author} {\bibfnamefont {N.}~\bibnamefont
  {Gr{\"u}n}},\ }\bibfield  {title} {\bibinfo {title} {Asymptotic behaviour of
  distorted-wave models for ionisation at relativistic energies},\ }\href
  {https://doi.org/10.1088/0953-4075/22/9/009} {\bibfield  {journal} {\bibinfo
  {journal} {J. Phys. B}\ }\textbf {\bibinfo {volume} {22}},\ \bibinfo {pages}
  {1357} (\bibinfo {year} {1989})}\BibitemShut {NoStop}%
\bibitem [{\citenamefont {Deco}\ \emph {et~al.}(1990)\citenamefont {Deco},
  \citenamefont {Momberger},\ and\ \citenamefont
  {Gr{\"u}n}}]{deco_k-shell_1990}%
  \BibitemOpen
  \bibfield  {author} {\bibinfo {author} {\bibfnamefont {G.}~\bibnamefont
  {Deco}}, \bibinfo {author} {\bibfnamefont {K.}~\bibnamefont {Momberger}},\
  and\ \bibinfo {author} {\bibfnamefont {N.}~\bibnamefont {Gr{\"u}n}},\
  }\bibfield  {title} {\bibinfo {title} {$K$-shell ionisation in heavy ion
  collisions at relativistic energies},\ }\href
  {https://doi.org/10.1088/0953-4075/23/12/017} {\bibfield  {journal} {\bibinfo
   {journal} {J. Phys. B}\ }\textbf {\bibinfo {volume} {23}},\ \bibinfo {pages}
  {2091} (\bibinfo {year} {1990})}\BibitemShut {NoStop}%
\bibitem [{\citenamefont {Hillenbrand}\ \emph {et~al.}(2015)\citenamefont
  {Hillenbrand}, \citenamefont {Hagmann}, \citenamefont {Jakubassa-Amundsen},
  \citenamefont {Monti}, \citenamefont {Bana{\'s}}, \citenamefont {Blumenhagen},
  \citenamefont {Brandau}, \citenamefont {Chen}, \citenamefont {Fainstein},
  \citenamefont {De~Filippo}, \citenamefont {Gumberidze}, \citenamefont {Guo},
  \citenamefont {Lestinsky}, \citenamefont {Litvinov}, \citenamefont
  {M{\"u}ller}, \citenamefont {Rivarola}, \citenamefont {Rothard},
  \citenamefont {Schippers}, \citenamefont {Sch{\"o}ffler}, \citenamefont
  {Spillmann}, \citenamefont {Trotsenko}, \citenamefont {Zhu},\ and\
  \citenamefont {St{\"o}hlker}}]{hillenbrand_electron-capture--continuum_2015}%
  \BibitemOpen
  \bibfield  {author} {\bibinfo {author} {\bibfnamefont {P.-M.}\ \bibnamefont
  {Hillenbrand}}, \bibinfo {author} {\bibfnamefont {S.}~\bibnamefont
  {Hagmann}}, \bibinfo {author} {\bibfnamefont {D.~H.}\ \bibnamefont
  {Jakubassa-Amundsen}}, \bibinfo {author} {\bibfnamefont {J.~M.}\ \bibnamefont
  {Monti}}, \bibinfo {author} {\bibfnamefont {D.}~\bibnamefont {Bana{\'s}}},
  \bibinfo {author} {\bibfnamefont {K.-H.}\ \bibnamefont {Blumenhagen}},
  \bibinfo {author} {\bibfnamefont {C.}~\bibnamefont {Brandau}}, \bibinfo
  {author} {\bibfnamefont {W.}~\bibnamefont {Chen}}, \bibinfo {author}
  {\bibfnamefont {P.~D.}\ \bibnamefont {Fainstein}}, \bibinfo {author}
  {\bibfnamefont {E.}~\bibnamefont {De~Filippo}},  \emph {et~al.},\ }\bibfield  {title}
  {\bibinfo {title} {Electron-capture-to-continuum cusp in {U$^{88+}$} + {N$_2$}
  collisions},\ }\href {https://doi.org/10.1103/PhysRevA.91.022705} {\bibfield
  {journal} {\bibinfo  {journal} {Phys. Rev. A}\ }\textbf {\bibinfo {volume}
  {91}},\ \bibinfo {pages} {022705} (\bibinfo {year} {2015})}\BibitemShut
  {NoStop}%
\bibitem [{\citenamefont {Shah}\ \emph {et~al.}(2003)\citenamefont {Shah},
  \citenamefont {McGrath}, \citenamefont {Illescas}, \citenamefont {Pons},
  \citenamefont {Riera}, \citenamefont {Luna}, \citenamefont {Crothers},
  \citenamefont {O{\textquoteright}Rourke},\ and\ \citenamefont
  {Gilbody}}]{shah_shifts_2003}%
  \BibitemOpen
  \bibfield  {author} {\bibinfo {author} {\bibfnamefont {M.~B.}\ \bibnamefont
  {Shah}}, \bibinfo {author} {\bibfnamefont {C.}~\bibnamefont {McGrath}},
  \bibinfo {author} {\bibfnamefont {C.}~\bibnamefont {Illescas}}, \bibinfo
  {author} {\bibfnamefont {B.}~\bibnamefont {Pons}}, \bibinfo {author}
  {\bibfnamefont {A.}~\bibnamefont {Riera}}, \bibinfo {author} {\bibfnamefont
  {H.}~\bibnamefont {Luna}}, \bibinfo {author} {\bibfnamefont {D.~S.~F.}\
  \bibnamefont {Crothers}}, \bibinfo {author} {\bibfnamefont {S.~F.~C.}\
  \bibnamefont {O{\textquoteright}Rourke}},\ and\ \bibinfo {author}
  {\bibfnamefont {H.~B.}\ \bibnamefont {Gilbody}},\ }\bibfield  {title}
  {\bibinfo {title} {Shifts in electron capture to the continuum at low
  collision energies: {Enhanced} role of target postcollision interactions},\
  }\href {https://doi.org/10.1103/PhysRevA.67.010704} {\bibfield  {journal}
  {\bibinfo  {journal} {Phys. Rev. A}\ }\textbf {\bibinfo {volume} {67}},\
  \bibinfo {pages} {010704(R)} (\bibinfo {year} {2003})}\BibitemShut {NoStop}%
\bibitem [{\citenamefont {Bhattacharya}\ \emph {et~al.}(2005)\citenamefont
  {Bhattacharya}, \citenamefont {Deb}, \citenamefont {Roy}, \citenamefont
  {Sahoo},\ and\ \citenamefont {Crothers}}]{bhattacharya_shifting_2005}%
  \BibitemOpen
  \bibfield  {author} {\bibinfo {author} {\bibfnamefont {S.}~\bibnamefont
  {Bhattacharya}}, \bibinfo {author} {\bibfnamefont {N.~C.}\ \bibnamefont
  {Deb}}, \bibinfo {author} {\bibfnamefont {K.}~\bibnamefont {Roy}}, \bibinfo
  {author} {\bibfnamefont {S.}~\bibnamefont {Sahoo}},\ and\ \bibinfo {author}
  {\bibfnamefont {D.~S.~F.}\ \bibnamefont {Crothers}},\ }\bibfield  {title}
  {\bibinfo {title} {Shifting of the electron-capture-to-the-continuum peak in
  proton-helium collisions at 10 and 20 {keV}},\ }\href
  {https://doi.org/10.1103/PhysRevA.71.012714} {\bibfield  {journal} {\bibinfo
  {journal} {Phys. Rev. A}\ }\textbf {\bibinfo {volume} {71}},\ \bibinfo
  {pages} {012714} (\bibinfo {year} {2005})}\BibitemShut {NoStop}%
\end{thebibliography}%

\end{document}